%% file: N=3_corrected_nov22.tex
\documentclass[11pt]{article}
\usepackage{jheppub}
\usepackage{amsmath, amsfonts, amssymb, array}
\usepackage{mathrsfs}
\usepackage{slashed}
\newcolumntype{C}[1]{>{\centering\let\newline\\\arraybackslash\hspace{0pt}}m{#1}}
\makeatletter
\g@addto@macro\bfseries{\boldmath}
\makeatother
\usepackage[english]{babel}
\usepackage[autostyle]{csquotes}
\usepackage{float}

\newcommand{\be}{\begin{equation}}
\newcommand{\ee}{\end{equation}}
\newcommand{\bea}{\begin{eqnarray}}
\newcommand{\eea}{\end{eqnarray}}
\newcommand{\nn}{\nonumber}

\usepackage{hyperref}

\def\nn{\nonumber}

\def\nn{\nonumber}

\newcommand{\ben}[1]{\begin{eqnarray}\label{#1} }
\newcommand{\een}{\end{eqnarray}}

\newcommand{\DD}{{\cal D}}

\newcommand{\thc}{\text{h.c.}}
\newcommand{\Dslash}{\not{\hbox{\kern-4pt $D$}}}
\newcommand{\pslash}{\not{\hbox{\kern-4pt $\partial$}}}
\newcommand{\Dcslash}{\not{\hbox{\kern-4pt $\DD$}}}

\usepackage{graphicx}
\usepackage{amssymb}
\usepackage{amsthm}
\usepackage{amsmath}
\usepackage{slashed}
\usepackage{color}
\usepackage{bbold}
\usepackage[normalem]{ulem}

\numberwithin{equation}{section}


\makeatletter
\g@addto@macro\bfseries{\boldmath}
\makeatother

\title{$N=3$ Conformal Supergravity in Four Dimensions}

\author{Subramanya Hegde$^{1}$,}
\author{Madhu Mishra$^{2}$}
\author{and Bindusar Sahoo$^{2}$}

\affiliation{$^1$ Harish Chandra Research Institute, HBNI, Jhunsi, Allahabad - 211019, India\\
	$^2$ Indian Institute of Science Education and Research,
	Vithura, Thiruvananthapuram - 695551, India}

\emailAdd{subramanyahegde@hri.res.in}
\emailAdd{madhu50315@iisertvm.ac.in}
\emailAdd{bsahoo@iisertvm.ac.in}

\abstract{In this paper we derive the action for $N=3$ conformal supergravity in four space-time dimensions. We construct a density formula for $N=3$ conformal supergravity based on the superform action principle. Finally, we embed the $N=3$ Weyl multiplet in the density formula to obtain the invariant action for $N=3$ conformal supergravity. There are two inequivalent embeddings by changing a particular coefficient from real to imaginary. They lead to invariant actions, which will either be the supersymmetrization of the Weyl square term or the Pontryagin density in the eventuality of gauge fixing to Poincar\'{e} supergravity. {As a consistency check of our formalism, we will show that the supersymmetrization of the Pontryagin density is a total derivative. We will demonstrate this for purely bosonic terms. We will also present the complete action for the supersymmetrization of Weyl square term.} We also discuss consistent truncation of $N=4$ Weyl multiplet to $N=3$ Weyl multiplet and use it for a robust check of our results using the earlier known results in $N=4$ conformal supergravity.
}

\begin{document}
	\allowdisplaybreaks
	\maketitle

	\section{Introduction}
	The study of conformal supergravity plays a vital role in constructing matter coupled supergravity theories with higher derivative corrections. The higher degree of symmetry in conformal supergravity allows one to arrange the degrees of freedom in shorter multiplets and the problem of constructing matter coupled theories in conformal supergravity becomes tractable. Upon using some of the multiplets as ``compensators'' to gauge fix the additional symmetries, one gets the physical matter-coupled supergravity theory with super-Poincar\'{e} symmetry. The crucial ingredients in this construction are the multiplets that form a representation of the underlying soft-superconformal algebra, which in the case of four dimensions is $SU(2,2|N)$. Here $N$-refers to the number of Majorana supercharges, and ``soft superconformal algebra'' is a modified superconformal algebra with field dependent structure constants. The most crucial multiplet that one encounters in conformal supergravity is the ``Weyl multiplet'' that contains the graviton and its supersymmetric partner gravitino along with a bunch of auxiliary fields needed to close the multiplet. Apart from the Weyl multiplet, one needs matter multiplets which perform a dual role.  They could play the role of ``compensators''  to gauge fix the additional symmetries of conformal supergravity and obtain supergravity with super-Poincar\'{e} symmetries. Such multiplets are called compensating multiplets and become a part of the supergravity multiplet after gauge fixing. The matter multiplets could also play the role of physical matter multiplets coupled to supergravity rather than playing the role of compensators.
	
	The process of constructing matter coupled supergravity theories with higher derivative corrections has been well studied for various extended supergravity theories up to six spacetime dimensions. The reason one cannot extend this approach beyond six dimensions is that there is no superconformal algebra beyond six dimensions \cite{Nahm:1977tg}. In four spacetime dimensions, the $N=2$ and $N=4$ Weyl multiplets were found in \cite{deWit:1979dzm} and \cite{Bergshoeff:1980is}, respectively. Another variant of the $N=2$ Weyl multiplet called the dilation Weyl multiplet was found recently in \cite{Butter:2017pbp} which was further used to construct supersymmetrization of arbitrary curvature squared invariants in $N=2$ supergravity \cite{Mishra:2020jlc}. The action for $N=2$ conformal supergravity was found in \cite{Bergshoeff:1980is} and the procedure of superconformal multiplet calculus to study matter coupled $N=2$ supergravity theories with or without higher derivative corrections is very well studied. Compared to $N=2$, the $N=4$ theories have been less explored in the superconformal formalism due to the lack of a conformal supergravity action until recently, when the most general $N=4$ conformal supergravity action was constructed \cite{Ciceri:2015qpa,Butter:2016mtk,Butter:2019edc}. However, compared to both $N=2$ and $N=4$ the $N=3$ theory is in a nascent stage. The most crucial component for $N=3$ conformal supergravity, i.e., the Weyl multiplet, was lacking for a long time. The components of the $N=3$ Weyl multiplet was predicted long back in \cite{FRADKIN1985233}. However, it did not give the complete supersymmetry transformation laws for the multiplet, and this gap was filled up recently in \cite{vanMuiden:2017qsh,Hegde:2018mxv}. 
	
	Upon the discovery of the $N=3$ Weyl multiplet, the most important aspect to study is the construction of the $N=3$ conformal supergravity action, which we plan to pursue in this paper. The construction of supergravity action is facilitated by the use of density formulae such as the chiral density formula \cite{deRoo:1980mm,Mohaupt:2000mj} or the tensor-vector density formula \cite{deWit:1982na,Claus:1997fk,deWit:2006gn} in $N=2$ theories. However, when one goes beyond $N=2$, there are no standard density formulae, and one needs to resort to some principle behind the construction of such density formulae, which can be generalized. Such a principle exists and goes by the name of  ``covariant superform action principle." Such a principle was first known in superspace as the ``ectoplasm principle'' \cite{Gates:1997ag,Gates:1997kr} and has been applied for the construction of the most general $N=4$ conformal supergravity actions in \cite{Butter:2016mtk,Butter:2019edc}. The method is very close to what is used in the rheonomy based approach to supergravity \cite{DAURIA1984423,Castellani:1991eu}. This method is not only useful to construct actions in theories with $N>2$, but it equally helps in constructing new density formulae and new invariant actions in $N=2$ conformal supergravity, which were not possible using the chiral or the tensor-vector density formula. For instance, the covariant superform approach was used in \cite{Hegde:2019ioy} to construct a new density formula in $N=2$ conformal supergravity which was further used to construct a new higher derivative action for the $N=2$ tensor multiplet. We will elaborate on this method in our paper.
	
	$N=3$ matter coupled supergravity has been studied using alternate approaches such as harmonic superspace or group manifold \cite{BRINK1978417,Galperin:1986id,CASTELLANI1986317} and has been of recent interest from the perspective of AdS-CFT correspondence \cite{Karndumri:2016miq,Karndumri:2016tpf}. However a full superconformal approach to the study of matter coupled $N=3$ supergravity is lacking and our current effort in this paper is to fill this gap. This paper only addresses the construction of $N=3$ conformal supergravity. Subsequently we would need to couple $N=3$ vector multiplets to conformal supergravity so that we can obtain matter coupled $N=3$ supergravity theories by using some of the vector multiplets as compensators and subsequently study their gaugings.
	
	This paper is organized as follows. In section-\ref{weyl}, we briefly describe the $N=3$ Weyl multiplet. In section-\ref{sec-density} we give the details of the covariant superform approach and use it to construct a density formula in $N=3$ conformal supergravity. In section-\ref{sec-actions} we discuss the embeding of $N=3$ Weyl multiplet in the density formula and present invariant actions.  We will show that there are two inequivalent choices of the embedding either with a real coefficient or an imaginary coefficient and the choices lead us to an invariant action which in the eventuality of gauge fixing to Poincar\'{e} supergravity would be the supersymmetrization of the Weyl square term or the Pontryagin density. This is not as simple as taking the real and imaginary part of the action in $N=2$ conformal supergravity constructed using chiral density formula. We will see that the density formula for $N=3$ just like its $N=4$ cousin mixes the chiral and anti-chiral part unlike the chiral density formula of $N=2$ which is an invariant action just for the chiral part and its conjugate is the invariant action for the anti-chiral part. {However, in the end we would expect that the supersymmetrization of the Pontryagin density is a total derivative. We would demonstrate this by giving the purely bosonic terms which can easily be seen to be a total derivative. This would act as a consistency check of our formalism. In this section, we will also give all the terms which arise in the supersymmetrization of the Weyl square term and are manifestly supercovariant.} In section-\ref{sec-reduction}, we discuss the off-shell reduction of the $N=4$ Weyl multiplet to the $N=3$ Weyl multiplet which provides a non-trivial consistency check for our results. In appendix-\ref{appendix-conventions}, we have explained the conventions for the $N=3$ superconformal algebra used in this paper and how it deviates from the conventions followed in \cite{vanMuiden:2017qsh,Hegde:2018mxv}. In appendix-\ref{appendix-Pontryagin} and appendix-\ref{appendix-weylsquare} we present all the relevant composites to obtain the actions given in section-\ref{sec-actions}.  In section-\ref{sec-conclusion}, we will end with some conclusions and future directions.
	\section{$N=3$ Weyl multiplet}\label{weyl}
	The $N=3$ Weyl multiplet is a $64+64$ (bosonic+fermionic) multiplet whose components are as tabulated in Table-\ref{Table-Weyl}.
	\begin{table}[t]
		\caption{Field content of the N = 3 Weyl multiplet}\label{Table-Weyl}
		\begin{center}
			\begin{tabular}{ | C{2cm}|C{2cm}|C{3cm}|C{2cm}|C{2cm}| }
				\hline
				Field & SU(3) Irreps & Restrictions &Weyl weight (w) & Chiral weight (c) \\ \hline
				$e_{\mu}{}^{a}$ & $\bf{1}$ & Vielbein & -1 & 0 \\ \hline
				$V_{\mu}{}^{i}{}_{j}$ & $\bf{8}$ & $(V_{\mu}{}^{i}{}_{j})^{*}\equiv V_{\mu}{}_{i}{}^{j}=-V_{\mu}{}^{j}{}_{i}$ SU(3)$_R$ gauge field &0 & 0  \\ \hline
				$A_{\mu}$ & $\bf{1}$ & U(1)$_R$ gauge field &0 & 0  \\ \hline
				$b_{\mu}$ & $\bf{1}$ & dilatation gauge field &0 & 0  \\ \hline
				$T^{i}_{ab}$ & $\bf{3}$ & Self-dual i.e $T^{i}_{ab}=\frac{1}{2}\varepsilon_{abcd}T^{i}{}^{cd}$ &1 & 1  \\ \hline
				$E_{i}$ &$\bf{\bar{3}}$ & Complex & 1&-1\\ \hline
				$D^{i}{}_{j}$ & $\bf{8}$ & $(D^{i}{}_{j})^{*}\equiv D_{i}{}^{j}=D^{j}{}_{i}$ &2 & 0  \\ \hline
				$\psi_{\mu}{}^{i}$ & $\bf{3}$ & $\gamma_{5}\psi_{\mu}{}^{i}=\psi_{\mu}{}^{i}$&-1/2 & -1/2  \\ \hline
				$\chi_{ij}$ & $\bf{\bar{6}}$ & $\gamma_{5}\chi_{ij}=\chi_{ij}$ &3/2 & -1/2  \\ \hline
				$\zeta^{i}$ & $\bf{3}$ & $\gamma_{5}\zeta^{i}=\zeta^{i}$ & 3/2 &-1/2 \\ \hline
				$\Lambda_{L}$ & $\bf{1}$ & $\gamma_{5}\Lambda_{L}=\Lambda_{L}$ &1/2 &-3/2 \\ \hline
			\end{tabular}
		\end{center}
	\end{table}
	In conformal supergravity, there are two kinds of supersymmetry which are labelled as Q (or ordinary) supersymmetry and S (or special) supersymmetry. The $Q$ and $S$-supersymmetry transformations of the components of the Weyl multiplet is given as:
	\begin{align}\label{N3susy}
	\delta e_{\mu}^{a}&= \bar{\epsilon}_{i}\gamma^{a}\psi_{\mu}^{i}+\thc \nonumber \\
	\delta \psi_{\mu}^{i}&=2\mathcal{D}_{\mu}\epsilon^{i}-\frac{1}{8}\varepsilon^{ijk}\gamma\cdot T_{j}\gamma_{\mu}\epsilon_{k}-\varepsilon^{ijk}\bar{\epsilon}_{j}\psi_{\mu k}\Lambda_{L}-\gamma_{\mu}\eta^{i} \nonumber \\
	\delta V_\mu{}^i{}_j &=\bar{\epsilon}^i\phi_{\mu j}- \frac{1}{48}\bar{\epsilon}^i\gamma_\mu\zeta_j+ \frac{1}{16}\varepsilon_{jkl}\bar{\epsilon}^k\gamma_\mu\chi^{il}- \frac{1}{16}\bar{\epsilon}^i\gamma\cdot T_j \gamma_\mu\Lambda_R- \frac{1}{16}\bar{\epsilon}^i\gamma_\mu \Lambda_R E_j +\frac{1}{8}\varepsilon_{klj}E^i\bar{\epsilon}^k\psi_\mu^l \nonumber \\
	&\quad+ \frac{1}{4}\bar{\epsilon}^i\gamma^a\psi_{\mu j}\bar{\Lambda}_L\gamma_a\Lambda_R-\bar{\psi}_\mu^i\eta_j-\thc-\text{trace} \nonumber \\
	\delta A_\mu &=\frac{i}{6}\bar{\epsilon}^i\phi_{\mu i}+ \frac{i}{36}\bar{\epsilon}^i\gamma_\mu\zeta_i+ \frac{i}{12}\varepsilon_{klp}E^p\bar{\epsilon}^k\psi_{\mu}^l+ \frac{i}{12}\bar{\epsilon}^i\gamma\cdot T_i\gamma_\mu\Lambda_R+\frac{i}{12}\bar{\epsilon}^i\gamma_\mu\Lambda_RE_i-\frac{i}{3}\bar{\epsilon}^i\gamma^a\psi_{\mu i}\bar{\Lambda}_L\gamma_a\Lambda_R\nonumber\\
	&\quad-\frac{i}{6}\bar{\psi}_\mu^i\eta_i+\thc \nonumber \\
	\delta b_\mu &= \frac{1}{2}(\bar{\epsilon}^i\phi_{\mu i}-\bar{\psi}_\mu^i\eta_i)+\thc\nonumber \\
	\delta \Lambda_L&=-\frac{1}{4}E_i\epsilon^i+\frac{1}{4}\gamma\cdot T_i\epsilon^i\nonumber \\
	\delta E_i &=-4 \bar{\epsilon}_i\slashed{D}\Lambda_L-\frac{1}{2}\varepsilon_{ijk}\bar{\epsilon}^j\zeta^k+\frac{1}{2}\bar{\epsilon}^j\chi_{ij}-\frac{1}{2}\varepsilon_{ijk}E^k\bar{\epsilon}^j\Lambda_L-4\bar{\Lambda}_L\Lambda_L\bar{\epsilon}_i\Lambda_R- 4\bar{\eta}_i\Lambda_L\nonumber \\
	\delta T^i_{ab} &= -\bar{\epsilon}^i\slashed{D}\gamma_{ab}\Lambda_R-4\varepsilon^{ijk}\bar{\epsilon}_jR_{ab}(Q)_k+\frac{1}{8}\bar{\epsilon}_j\gamma_{ab}\chi^{ij}+\frac{1}{24}\varepsilon^{ijk}\bar{\epsilon}_j\gamma_{ab}\zeta_k-\frac{1}{8}\varepsilon^{ijk}E_j\bar{\epsilon}_k\gamma_{ab}\Lambda_R\nonumber \\
	&\quad+\bar{\eta}^i\gamma_{ab}\Lambda_R\nonumber \\
	\delta \chi_{ij}&=2\slashed{D}E_{(i}\epsilon_{j)}-8\varepsilon_{kl(i}\gamma\cdot R(V)^l{}_{j)}\epsilon^k-2\gamma\cdot\slashed{D}T_{(i}\epsilon_{j)}+\frac{1}{3}\varepsilon_{kl(i}D^l{}_{j)}\epsilon^k\nonumber \\&\quad+\frac{1}{4}\varepsilon_{kl(i}E^k\gamma\cdot T_{j)}\epsilon^l-\frac{1}{3}\bar{\Lambda
	}_L\gamma_a\epsilon_{(i}\gamma^a\zeta_{j)}+\frac{1}{4}\varepsilon_{lm(i}E_{j)}E^m\epsilon^l-\bar{\Lambda}_L\gamma^a\Lambda_R\gamma_aE_{(i}\epsilon_{j)}\nonumber\\
	&\quad-\bar{\Lambda}_L\gamma\cdot T_{(i}\gamma^a\Lambda_R\gamma_a\epsilon_{j)}+ 2\gamma\cdot T_{(i}\eta_{j)}+ 2E_{(i}\eta_{j)}\nonumber \\
	\delta \zeta^i &=- 3\varepsilon^{ijk}\slashed{D}E_j\epsilon_k +\varepsilon^{ijk}\gamma\cdot\slashed{D}T_k\epsilon_j-4\gamma\cdot R(V)^i{}_j\epsilon^j-16i\gamma\cdot R(A)\epsilon^i-\frac{1}{2}D^i{}_j\epsilon^j-\frac{3}{8}E^i\gamma\cdot T_j\epsilon^j\nonumber\\
	&\quad+\frac{3}{8}E^j\gamma\cdot T_j\epsilon^i+\frac{3}{8}E^iE_j\epsilon^j+\frac{1}{8}E^jE_j\epsilon^i\nonumber\\
	&\quad- 4 \bar{\Lambda}_L\slashed{D}\Lambda_{R}\epsilon^i- 4 \bar{\Lambda}_R\slashed{D}\Lambda_L\epsilon^i- 3\bar{\Lambda}_R\slashed{D}\gamma_{ab}\Lambda_L\gamma^{ab}\epsilon^i-3\bar{\Lambda}_L\gamma_{ab}\slashed{D}\Lambda_R\gamma^{ab}\epsilon^i\nonumber\\
	&\quad+\frac{1}{2}\varepsilon^{ijk}\bar{\Lambda}_L\gamma^a\epsilon_j\gamma_a\zeta_k-6\bar{\Lambda}_L\Lambda_L\bar{\Lambda}_R\Lambda_R\epsilon^i+\varepsilon^{ijk}\gamma\cdot T_j\eta_k-3\varepsilon^{ijk}E_j\eta_k\nonumber\\
	\delta D^i_j&=-3\bar{\epsilon}^i\slashed{D}\zeta_j-3\varepsilon_{jkl}\bar{\epsilon}^k\slashed{D}\chi^{il}+\frac{1}{4}\varepsilon_{jkl}\bar{\epsilon}^i\zeta^k E^l+\frac{1}{2}\varepsilon_{jkl}\bar{\epsilon}^k\zeta^l E^i+\frac{3}{4}\bar{\epsilon}^i\chi_{jk}E^k+ 3\bar{\epsilon}^i\gamma\cdot T_j\overset{\leftrightarrow}{\slashed{D}}\Lambda_R\nonumber\\
	&\quad-3\bar{\epsilon}^i\slashed{D}\Lambda_RE_j-3\bar{\epsilon}^i\slashed{D}E_j\Lambda_R+ \frac{3}{4}\varepsilon_{jkl}E^l\bar{\epsilon}^k\Lambda_LE^i+ {3\varepsilon_{jkl}T^i\cdot T^l\bar{\epsilon}^k\Lambda_{L}}-2\bar{\epsilon}^i\Lambda_L\bar{\Lambda}_R\zeta_j\nonumber\\
	&\quad-3\bar{\epsilon}^i\Lambda_L\bar{\Lambda}_R\Lambda_RE_j+3\bar{\epsilon}^i\gamma\cdot T_j\Lambda_L\bar{\Lambda}_R\Lambda_R+\thc-\text{trace}
	\end{align}
	where, $\mathcal{D}_{\mu}\epsilon^{i}$ is defined as\footnote{Our conventions for the superconformal transformations are explained in appendix-\ref{appendix-conventions}.}:
	\begin{align}\label{Depsilon}
	\mathcal{D}_\mu\epsilon^i=\partial_\mu\epsilon^i-\frac{1}{4}\gamma\cdot\omega_\mu\epsilon^i+\frac{1}{2}(b_\mu+iA_\mu)\epsilon^i-V_\mu{}^i{}_j\epsilon^j 
	\end{align}
	We also give the complete Q and S-supersymmetry transformations of the dependent gauge fields corresponding to local Lorentz transformations ($\omega_{\mu}^{ab}$), S-supersymmetry ($\phi_{\mu}^{i}$) and special conformal transformation or K-gauge field ($f_{\mu}^{a}$).
	\begin{align}\label{deptransf}
	\delta \omega_\mu^{ab}&=-\frac{1}{2}\bar{\epsilon}^i\gamma^{ab}\phi_{\mu i}+\frac{1}{2}\varepsilon_{ijk}\bar{\epsilon}^i\psi_\mu^jT^{ab k}+\bar{\epsilon}^i\gamma_\mu R(Q)^{ab}{}_i-\frac{1}{2}\bar{\eta}^i\gamma^{ab}\psi_{\mu i}+\thc\nonumber\\
	\delta \phi_\mu^i&= -\frac{i}{12}(\gamma_\mu\gamma\cdot R(A)-3\gamma\cdot R(A)\gamma_\mu)\epsilon^i-\frac{1}{6}(3\gamma\cdot R(V){}^i{}_j\gamma_\mu-\gamma_\mu\gamma\cdot R(V){}^i{}_j)\epsilon^j\nonumber\\
	&\quad-\frac{1}{4}\varepsilon^{ijk}\bar{\Lambda}_L\gamma_{\mu}R_{ab}(Q)_k\gamma^{ab}\epsilon_j+ \frac{1}{32}\gamma\cdot T^{[i}\gamma_{\mu}\gamma\cdot T_j\epsilon^{j]}+\frac{1}{24}\varepsilon^{ijk}(\gamma_{\mu}\gamma\cdot\slashed{D}T_j-3\slashed{D}\gamma\cdot T_j\gamma_\mu)\epsilon_k\nonumber\\
	& \quad -\frac{1}{12}\bar{\epsilon}^{[i}\psi_{\mu}^{k]}\zeta_{k}+\frac{1}{32}\Big(\bar{\epsilon}^{i}\gamma_{a}\psi_{\mu j}-\delta^{i}_{j}\bar{\epsilon}^{k}\gamma_{a}\psi_{\mu k}+\bar{\epsilon}_{j}\gamma_{a}\psi_{\mu}^{i}-\delta^{i}_{j}\bar{\epsilon}_{k}\gamma_{a}\psi_{\mu}^{k}\Big)\gamma^{a}\Lambda_{L}E^{j} \nonumber \\
	&\quad +\frac{1}{96}\Big(\bar{\epsilon}^{i}\gamma_{a}\psi_{\mu j}-\delta^{i}_{j}\bar{\epsilon}^{k}\gamma_{a}\psi_{\mu k}+\bar{\epsilon}_{j}\gamma_{a}\psi_{\mu}^{i}-\delta^{i}_{j}\bar{\epsilon}_{k}\gamma_{a}\psi_{\mu}^{k}\Big)\gamma^{a}\zeta^{j}-\frac{1}{4}\bar{\epsilon}^{[i}\psi_{\mu}^{j]}E_{j}\Lambda_R\nonumber \\
	&\quad  -\frac{1}{32}\varepsilon^{ijk}\Big(\bar{\epsilon}^{l}\gamma_{a}\psi_{\mu k}+\bar{\epsilon}_{k}\gamma_{a}\psi_{\mu}^{l}\Big)\gamma^{a}\chi_{jl}-\frac{1}{8}\varepsilon_{jkl}\bar{\epsilon}^{j}\psi_{\mu}^{k}\chi^{il}-\frac{1}{2}\varepsilon^{ijk}\bar{\epsilon}_{j}\psi_{\mu k}\slashed{D}\Lambda_{L}\nonumber \\
	&\quad -\frac{1}{16}\Big(\bar{\epsilon}^{i}\gamma_{a}\psi_{\mu j}-\delta^{i}_{j}\bar{\epsilon}^{k}\gamma_{a}\psi_{\mu k}+\bar{\epsilon}_{j}\gamma_{a}\psi_{\mu}^{i}-\delta^{i}_{j}\bar{\epsilon}_{k}\gamma_{a}\psi_{\mu}^{k}\Big)\gamma\cdot T^{j}\gamma^{a}\Lambda_{L}-\frac{1}{4}\varepsilon^{ijk}\bar{\epsilon}_{j}\gamma_{a}\phi_{\mu k}\gamma^{a}\Lambda_{L}\nonumber \\
	&\quad+2\mathcal{D}_{\mu}\eta^{i}-\frac{1}{24}\varepsilon^{ijk}\gamma_\mu\gamma\cdot T_j\eta_k+\frac{1}{4}\varepsilon^{ijk}\bar{\eta}_{j}\gamma_{a}\psi_{\mu k}\gamma^{a}\Lambda_{L}\nonumber \\
	\delta f_\mu{}^a&=-\bar{\epsilon}^i\gamma_\mu D_b R(Q)^{ab}{}_i+\frac{1}{4}\bar{\epsilon}^i {\tilde{R}(S)}^a{}_{\mu i}+\frac{1}{2}\varepsilon_{ijk}T_{\mu b}{}^k\bar{\epsilon}^iR(Q)^{abj}-\frac{i}{6}\bar{\epsilon}^{k}\gamma_{b}\psi_{\mu k}\tilde{R}(A)^{ab}\nonumber \\
	& \quad +\frac{1}{3}\bar{\epsilon}^{i}\gamma_{b}\psi_{\mu j}\tilde{R}(V)^{ab}{}^{j}{}_{i}+\frac{1}{64}\bar{\epsilon}^{[i}\gamma\cdot T_{i}\gamma^{a}\gamma\cdot T^{j]}\psi_{\mu j}-\frac{1}{3}\varepsilon_{ijk}\bar{\epsilon}^{i}\psi_{\mu}^{j}D_{b}T^{ab k}+\frac{1}{48}\varepsilon_{ijk}\bar{\epsilon}^{i}\gamma^{a}\gamma\cdot T^{j}\phi_{\mu}^{k}\nonumber \\
	&\quad +\frac{1}{2}\bar{\eta}^{i}\gamma^{a}\phi_{\mu i} -\frac{1}{4}\bar{\eta}^{i}R(Q)_{\mu}{}^{a}{}_{i}-\frac{1}{48}\varepsilon_{ijk}\bar{\eta}^{i}\gamma\cdot T^{k}\gamma^{a}\psi_{\mu}^{j}+\thc,\nonumber\\
	\end{align}
	The above mentioned supersymmetry transformations of the dependent gauge fields can be obtained from the independent ones as explained in \cite{Freedman:2012zz}. We can also obtain the supersymmetry transformations (Q as well as S) of the curvatures as shown below:
	\begin{align}\label{curvtransf}
	\delta R(M)_{abcd}&=\frac{1}{4}\bar{\epsilon}^i\gamma_{ab}R(S)^{-}_{cdi}+\frac{1}{4}\bar{\epsilon}^i\gamma_{cd}R(S)^{-}_{abi}-\frac{1}{4}\bar{\epsilon}^i\slashed{D}\gamma_{ab}R(Q)_{cdi}-\frac{1}{4}\bar{\epsilon}^i\slashed{D}\gamma_{cd}R(Q)_{abi}\nonumber \\
	&\quad +\frac{3}{4}\bar{\eta}^{i}\gamma_{ab}R(Q)_{cd i}+\frac{3}{4}\bar{\eta}^{i}\gamma_{cd}R(Q)_{ab i}+\thc\nonumber\\
	\delta R(S)^{+}_{ab}{}^{i}&= -2D^{c}R(M)^{+}_{abcd}\gamma^d \epsilon^{i}-\frac{1}{4}\left(\gamma^{cd}\gamma_{ab}+\frac{1}{3}\gamma_{ab}\gamma^{cd}\right)\Big(-\slashed{D}R(V)_{cd}{}^{i}{}_j\epsilon^{j}+\frac{i}{2}\slashed{D} R(A)_{cd}\epsilon^{i}\nonumber \\
	& -\frac{1}{2}D_{f}T^{fe}_{j}T_{cd}^{[j}\gamma_e\epsilon^{i]}-\frac{1}{4}T^{fe}_{j}D_{f}T_{cd}^{[j}\gamma_{e}\epsilon^{i]}+\epsilon_{j}D_{c}D^{e}T_{edk}\varepsilon^{ijk}-\frac{1}{16}\epsilon_{l}T_{cd}^{[p}T_{p}\cdot T_{q}\varepsilon^{i]lq}\Big)\nonumber \\
	& +\frac{1}{8}\varepsilon^{ijk}\bar{\epsilon}_{j}R(Q)_{cd k}\gamma^{cd}\gamma_{ab}\slashed{D}\Lambda_{L}-\frac{1}{2}\varepsilon^{ijk}\bar{\epsilon}_{j}R(Q)_{ab k}\slashed{D}\Lambda_{L}+\frac{1}{8}\varepsilon^{ijk}\bar{\epsilon}_{j}D_{e}R(Q)_{cd k}\gamma^{cd}\gamma_{ab}\gamma^{e}\Lambda_{L}\nonumber \\
	&-\frac{1}{4}\varepsilon^{ijk}\bar{\epsilon}_{j}\gamma_{c}\slashed{D}R(Q)_{abk}\gamma^{c}\Lambda_{L}+\frac{1}{96}\bar{\epsilon}^{[i}\gamma^{f}R(Q)_{cdk}\gamma_{ab}\gamma_{cd}\gamma\cdot T^{k]}\gamma_{f}\Lambda_{L}\nonumber \\
	&+\frac{1}{32}\bar{\epsilon}^{[i}\gamma^{f}R(Q)_{cdk}\gamma_{cd}\gamma_{ab}\gamma\cdot T^{k]}\gamma_{f}\Lambda_{L}+\frac{1}{16}\bar{\epsilon}^{[i}\gamma^{c}R(Q)_{abj}\gamma_{c}\Lambda_{L}E^{j]}+\frac{1}{48}\bar{\epsilon}^{[i}\gamma^{c}R(Q)_{abj}\gamma_{c}\zeta^{j]}\nonumber \\
	&-\frac{1}{32}\varepsilon^{ijk}\bar{\epsilon}^{l}\gamma_{c}R(Q)_{abk}\gamma^{c}\chi_{jl}-\frac{3}{4}\left(\gamma^{cd}\gamma_{ab}+\frac{1}{3}\gamma_{ab}\gamma^{cd}\right)\left(\eta^{j}R(V)_{cd j}{}^{i}+\frac{i}{2}\eta^{i}R(A)_{cd}\right)\nonumber \\
	&\quad +\frac{1}{2}R(M)_{abcd}\gamma^{cd}\eta^{i}\nonumber \\
	\delta R(Q)_{ab}^{i}&=-\frac{1}{2}R(M)_{abcd}\gamma^{cd}\epsilon^{i}+\frac{1}{4}\Big[\gamma^{cd}\gamma_{ab}+\frac{1}{3}\gamma_{ab}\gamma^{cd}\Big]\Big[R(V)_{cd}{}^{i}{}_j\epsilon^{j}-\frac{i}{2}R(A)_{cd}\epsilon^{i}-\frac{1}{4}\varepsilon^{ijk}\slashed{D}T_{cd k}\epsilon_{j}\Big] \nonumber \\
	&\quad -\frac{1}{8}\varepsilon^{ijk}\left[\gamma\cdot T_{k}\gamma_{ab}+\frac{1}{3}\gamma_{ab}\gamma\cdot T_{k}\right]\eta_{j}\nonumber \\
	\delta R(A)_{ab}&=-\frac{i}{18}\bar{\epsilon}^{k}\gamma_{[a}D_{b]}\zeta_{k}+\frac{i}{6}\bar{\epsilon}^{k}R(S)_{abk}+\frac{i}{6}\bar{\epsilon}^{j}D_{[a}\left(\gamma\cdot T_{j}\gamma_{b]}\Lambda_{R}\right)-\frac{i}{6}\bar{\epsilon}^{j}\gamma_{[a}D_{b]}\left(\Lambda_{R}E_{j}\right)\nonumber \\
	&\quad +\frac{i}{12}\varepsilon_{ijk}\bar{\epsilon}^{i}R(Q)_{ab}^{j}E^{k}-\frac{i}{3}\bar{\epsilon}^{k}\gamma^{c}R(Q)_{abk}\bar{\Lambda}_{L}\gamma_{c}\Lambda_{R}-\frac{i}{72}\varepsilon_{ijk}\bar{\epsilon}^{i}\zeta^{j}T_{ab}^{k}\nonumber \\
	&\quad +\frac{i}{96}\varepsilon_{ijk}\bar{\epsilon}^{i}\gamma_{[a}\gamma\cdot T^{j}\gamma\cdot T^{k}\gamma_{b]}\Lambda_{L}+\frac{i}{24}\varepsilon_{ijk}\bar{\epsilon}^{i}\Lambda_{L}T_{ab}^{j}E^{k}-\frac{i}{36}\bar{\eta}^{k}\gamma_{ab}\zeta_{k}+\frac{i}{6}\bar{\eta}^{k}R(Q)_{abk}\nonumber \\
	&\quad -\frac{i}{12}\bar{\eta}^{k}\gamma_{ab}\Lambda_{R}E_{k}-\frac{i}{3}\bar{\eta}^{k}\Lambda_R T_{abk}+\text{h.c}\nonumber \\
	\delta R(V)_{ab}{}^{i}{}_{j}&=\bar{\epsilon}^{i}R(S)_{abj}-\frac{1}{8}\varepsilon_{jkm}\bar{\epsilon}^{k}\gamma_{[a}D_{b]}\chi^{im}+\frac{1}{24}\bar{\epsilon}^{i}\gamma_{[a}D_{b]}\zeta_{j}+\frac{1}{8}\bar{\epsilon}^{i}\gamma_{[a}D_{b]}\left(E_{j}\Lambda_{R}\right)-\frac{1}{16}\bar{\epsilon}^{[i}\chi_{jm}T_{ab}^{m]}\nonumber \\
	&\quad -\frac{1}{96}\varepsilon_{jlm}\bar{\epsilon}^{l}\zeta^{i}T_{ab}^{m}-\frac{1}{32}\varepsilon_{jlm}\bar{\epsilon}^{l}\Lambda_{L}E^{j}T_{ab}^{m}-\frac{1}{128}\varepsilon_{jkl}\bar{\epsilon}^{k}\gamma_{[a}\gamma\cdot T^{l}\gamma\cdot T^{i}\gamma_{b]}\Lambda_{L}\nonumber \\
	&\quad +\frac{1}{8}E^{i}\varepsilon_{jkl}\bar{\epsilon}^{k}R(Q)_{ab}^{l}-\frac{1}{8}\bar{\epsilon}^{i}D_{[a}\left(\gamma\cdot T_{j}\gamma_{b]}\Lambda_{R}\right)+\frac{1}{4}\bar{\epsilon}^{i}\gamma^{c}R(Q)_{abj}\bar{\Lambda}_{L}\gamma_{c}\Lambda_{R}\nonumber \\
	&\quad +\bar{\eta}^{i}R(Q)_{abj}+\frac{1}{4}T_{abj}\bar{\eta}^{i}\Lambda_{R}-\frac{1}{16}\varepsilon_{jkl}\bar{\eta}^{k}\gamma_{ab}\chi^{il}+\frac{1}{48}\bar{\eta}^{i}\gamma_{ab}\zeta_{j}\nonumber\\
	&\quad +\frac{1}{16}\bar{\eta}^{i}\gamma_{ab}\Lambda_{R}E_{j}-\left(\text{h.c;traceless}\right)
	\end{align}
	We can obtain the Bianchi identities satisfied by the curvatures as shown below:
	\begin{align}\label{Bianchi}
	R(D)_{ab}&=0 \nonumber \\
	R(M)_{abcd}&=R(M)_{cdab}\nonumber\\
	\varepsilon^{aecd}R(M)_{cdeb}&=0\nonumber \\
	\frac{1}{4}\varepsilon^{abcd}\varepsilon^{efgh}R(M)_{abef}&=R(M)^{cdgh}\nonumber\\
	\varepsilon^{cdef}D_{b}D_{d}R(M)_{efab}&=0 \nonumber \\
	R(K)_{ab}{}^{c}&=D_{e}R(M)_{ab}{}^{ec}\nonumber \\
	\varepsilon^{abcd}D_{b}R(V)_{cd}{}^{i}{}_{j}{}&=-\frac{1}{16}\bar{\Lambda}_{L}\gamma_{b}\gamma\cdot T^{i}R(Q)^{ab}_{j}+\text{(h.c; traceless)}\nonumber \\
	\varepsilon^{abcd}D_{b}R(A)_{cd}&=\frac{i}{12}\bar{\Lambda}_{L}\gamma_{b}\gamma\cdot T^{j}R(Q)^{ab}{}_{j}+\frac{i}{12}\bar{\Lambda}_{R}\gamma_{b}\gamma\cdot T_{j}R(Q)^{ab}{}^{j}\nonumber \\
	D_{a}R(Q)^{abi}&=-\frac{1}{4}\varepsilon^{abcd}\gamma_{a}R(S)_{cd}^{i}\nonumber \\
	R(Q)_{ab}^{+i}&=0 \nonumber \\
	R(S)_{ab}^{-i}&=\slashed{D}R(Q)_{ab}^{i}\nonumber \\
	\gamma^{ab}R(S)_{ab}^{i}&=0\nonumber \\
	\gamma^{a}R(S)_{ab}^{+i}&=0 \nonumber \\
	\varepsilon^{abcd}D_{b}R(S)_{cd}^{i}&=\frac{1}{12}\varepsilon^{ijk}\gamma^{a}T_{k}\cdot R(S)_{j}+\frac{1}{3}\varepsilon^{ijk}T^{ab}_{k}D^{d}R(Q)_{dbj}-\frac{1}{3}\gamma^{a}R(V)^{i}{}_{j}\cdot R(Q)^{j}\nonumber \\
	&\quad +\frac{i}{6}\gamma^{a}R(A)\cdot R(Q)^{i}-\frac{1}{3}\varepsilon^{ijk}D^{g}T_{gck}R(Q)^{ac}_{j}+\frac{1}{32}\gamma\cdot T^{[l}\gamma^{a}T_{l}\cdot R(Q)^{i]}
	\end{align}
	\section{Covariant superform approach and the $N=3$ density formula}\label{sec-density}
In this section we will briefly review the covariant superform approach that we will use to construct the $N=3$ density formula. This density formula will be further used to construct the $N=3$ conformal supergravity action. The discussion on the covariant superform approach closely follows that of \cite{Butter:2019edc,Hegde:2019ioy}. The basic idea behind the covariant superform approach is the following. Consider a $D$ dimensional manifold $\mathcal{N}$ and a d-dimensional submanifold $\mathcal{M}\subset \mathcal{N}$. Consider a d-form $J$ and an action-integral on $\mathcal{M}$ given as:
	\begin{align}\label{3.1}
	S=\int_{\mathcal{M}}J
	\end{align}
	Under arbitrary diffeomorphism ($\xi$) on the larger manifold $\mathcal{N}$, the d-form $J$ transforms as a Lie derivative which can be written as an anti-commutator of interior product and exterior derivative as shown below:
	\begin{align}\label{3.2}
	\delta_{\xi}J=\mathcal{L}_{\xi}J=d(i_{\xi}J)+i_{\xi}dJ
	\end{align}
	Hence, if we consider the variation of the action integral (\ref{3.1}) under arbitrary diffeomorphism on the larger manifold $\mathcal{N}$, it becomes invariant if the d-form $J$ is closed i.e.
	\begin{align}\label{3.3}
	dJ=0
	\end{align} 
	Now we can consider the following situation where the submanifold $\mathcal{M}$ is our spacetime manifold and the geometrization of some of the gauge symmetries leads us to the larger manifold $\mathcal{N}$. And hence in this geometric picture, the condition (\ref{3.3}) for the invariance of the action integral under arbitrary diffeomorphism on the larger manifold is the same as that for gauge invariance because now gauge symmetry becomes a part of an arbitrary diffeomorphism on the manifold $\mathcal{N}$. We will demonstrate this in particular for $N=3$ conformal supergravity where we will geometrize ordinary supersymmetry (or Q-supersummetry) to construct a larger manifold $\mathcal{N}$ containing the spacetime manifold $\mathcal{M}$.
	
Let us label the local coordinates on the manifold $\mathcal{N}$ by $Z^{M}=(x^{\mu},\theta^{\mathbb{m}})$, where $x^{\mu}$ are the coordinates on the spacetime manifold and $\theta^{\mathbb{m}}$ are fermionic coordinates required to geometrize Q-supersymmetry. The spacetime manifold $\mathcal{M}$ is obtained from $\mathcal{N}$ by the following truncation:
	\begin{align}\label{3.4}
	\mathcal{M}=\mathcal{N}_{|\theta=d\theta=0}
	\end{align}
The $\theta=0$ condition truncates the elements of $\mathcal{N}$ to that of the spacetime manifold while $d\theta=0$ condition truncates the elements of the tangent/cotangent space associated with every point of $\mathcal{N}$ to that of the spacetime manifold. Let us further decompose the Supervielbein 1-form on $\mathcal{N}$ as follows:
	\begin{align}\label{3.5}
	E^{A}=(e^a,\psi^{i}, \psi_{i})
	\end{align}
	When truncated to the spacetime manifold, the one-forms $e^a$, $\psi^{i}$ and $\psi_{i}$ are the standard vielbein one-form, left chiral and right chiral gravitino one-forms respectively\footnote{We will be working in the Dirac 4-component notation for the spinors.}. Let us construct the 4-form $J$ as:
	\begin{align}\label{3.6}
	J=J_{DCBA}E^AE^BE^CE^D
	\end{align}
where we have suppressed the wedge products between the 1-forms $E^A$. We also assume that the $J_{DCBA}$ is fully supercovariant. The action integral (\ref{3.1}) will be invariant under Q-supersymmetry if we ensure that $dJ=0$. However in $N=3$ conformal supergravity there are other gauge symmetries apart from Q-supersymmetry such as local Lorentz transformations, dilatations, $SU(3)\times U(1)$ R-symmetry, special conformal transformations and special (S)-supersymmetry. All these gauge transformations are easier to implement and hence we demand that by construction the 4-form $J$ is invariant under these gauge transformations. Hence the condition $dJ=0$ for invariance under Q-supersymmetry can be replaced by the condition for covariant closure
	\begin{align}\label{3.7}
	\nabla J=0
	\end{align}
	where $\nabla$ is covariant with respect to all the other gauge transformations mentioned above apart from Q-supersymmetry. We do this because the condition for covariant closure is easier to implement in a systematic way than the simple closure condition. In order to implement (\ref{3.7}), we need to know how the covariant exterior derivative acts on the supervielbeins and fully supercovariant objects. It acts on the supervielbein 1-form as $\nabla E^A=T^A$ to give the supertorsion 2-form $T^A$, the form of which is known as shown below due to our understanding of the $N=3$ Weyl multiplet.
	\begin{align}\label{3.8}
	\nabla e^{a} &\equiv t_{0} e^{a} \nonumber \\
	\nabla \psi^{i} &\equiv t_{3/2}\psi^{i}+t_{1}\psi^{i}+t_{1/2}\psi^{i}\nonumber \\
	\nabla \psi_{i} &\equiv \bar{t}_{3/2}\psi_{i}+\bar{t}_{1}\psi_{i}+\bar{t}_{1/2}\psi_{i}
	\end{align}
	where the operators $t_{k}$ are defined as\footnote{The subscripts on the $t_k$ denotes the Weyl weights of the supercovariant objects that appear in their expressions}:
	\begin{align}\label{3.9}
	t_{0} e^{a} &=-\frac{1}{2}\bar{\psi}^{i}\gamma^{a}\psi_{i}\equiv T^{a}\nonumber \\
	t_{3/2}\psi^{i}&=\frac{1}{2}e^{a}e^{b}R(Q)_{ba}{}^{i}\nonumber \\
	t_{1}\psi^{i}&=-\frac{1}{16}\varepsilon^{ijk}\gamma\cdot T_{j}e^{a}\gamma_{a}\psi_{k}\nonumber \\
	t_{1/2}\psi^{i}&=\frac{1}{4}\varepsilon^{ijk}\bar{\psi}_{j}\psi_{k}\Lambda_{L} \nonumber \\
	\bar{t}_{3/2}\psi_{i}&=\frac{1}{2}e^{a}e^{b}R(Q)_{ba}{}_{i}\nonumber \\
	\bar{t}_{1}\psi_{i}&=-\frac{1}{16}\varepsilon_{ijk}\gamma\cdot T^{j}e^{a}\gamma_{a}\psi^{k}\nonumber \\
	\bar{t}_{1/2}\psi_{i}&=\frac{1}{4}\varepsilon_{ijk}\bar{\psi}^{j}\psi^{k}\Lambda_{R}
	\end{align}
	The covariant exterior derivative will act on super covariant objects $\Phi$ as:
	\begin{align}\label{3.10}
	\nabla \Phi \equiv \left(\nabla_{1}+\nabla_{1/2}+\bar{\nabla}_{1/2}\right)\Phi
	\end{align}
	where,
	\begin{align}\label{3.11}
	\nabla_{1}\Phi&=e^{a}D_{a}\Phi\nonumber \\
	\nabla_{1/2}\Phi&=\frac{1}{2}\bar{\psi}^{k}\nabla_{k}\Phi \nonumber \\
	\bar{\nabla}_{1/2}\Phi&=\frac{1}{2}\bar{\psi}_{k}\nabla^{k}\Phi \nonumber \\
	\end{align}
	Here, $D_{a}$ is the fully supercovariant derivative whereas $\nabla_{k}$ and $\nabla^k$ are superspace covariant derivatives along the fermionic direction which play the role of supersymmetry generators from the spacetime perspective. Hence, when truncated to the spacetime manifold one may treat $\nabla_{k}$ as $\nabla^k$ as shown below:
	\begin{align}\label{3.12}
	\frac{1}{2} \bar{\psi}^{k}\nabla_{k}\Phi &=\delta^{Q}_{L}\left(\frac{1}{2}\psi^{k}\right)\Phi\nonumber \\
	\frac{1}{2} \bar{\psi}_{k}\nabla^{k}\Phi &=\delta^{Q}_{R}\left(\frac{1}{2}\psi_{k}\right)\Phi
	\end{align}
	where the R.H.S of the above equations means that you take the left and right supersymmetry transformations of the supercovariant object $\Phi$ and replace the transformation parameters by the appropriate gravitino 1-forms as mentioned in the bracket\footnote{Note that fully supercovariant derivative $D_{\mu}$ is related to the covariant derivative $\nabla_{\mu}$ by addition of explicit Q-covariantization terms i.e $D_{\mu}=\nabla_{\mu}-\delta^{Q}_{L}\left(\frac{1}{2}\psi^{k}\right)-\delta^{Q}_{R}\left(\frac{1}{2}\psi_{k}\right)$. Hence, when truncated to the spacetime manifold, eq-\ref{3.10} is consistent with the fact that $\nabla$ is covariant with respect to all the gauge transformations except Q-supersymmetry.}. Now we are in a position to implement the covariant closure condition (\ref{3.7})\footnote{We will follow the convention that the exterior derivative acts from the right.}. We need to have an ansatz for the block $J_{DCBA}$ associated with the highest number of gravitino one-forms. We choose the ansatz to be the following:
	\begin{align}\label{3.13}
	J_{\psi^{4}}&=\frac{1}{4}\bar{\psi}^{i}\psi^{j}\bar{\psi}^{k}\psi^{l}\varepsilon_{ijm}\varepsilon_{kln}C^{mn}\nonumber \\
	J_{\bar{\psi}^{4}}&=\frac{1}{4}\bar{\psi}_{i}\psi_{j}\bar{\psi}_{k}\psi_{l}\varepsilon^{ijm}\varepsilon^{kln}C_{mn}\;,
	\end{align}
	where $C^{mn}$ is in the $\bf{6}$ of SU(3) and its complex conjugate $C_{mn}$ is in the $\bf{\bar{6}}$ of SU(3). Further, $C^{ij}$ needs to have Weyl as well as chiral weights to be $+2$ and should be invariant under S-supersymmetry. Now we would need to impose the covariant closure condition (\ref{3.7}). We can schematically decompose the 5-form $\nabla J$ by the number of left chiral gravitino 1-forms ($\psi$), right chiral gravitino 1-forms ($\bar{\psi}$) and vielbein 1-forms ($e$) they carry as shown below (Note that in the equation below $\bar{\psi}$ should not be confused with the Dirac conjugate. The notation is used to schematically represent a right chiral gravitino).
	\begin{align}\label{3.14}
	\nabla J=\sum_{m,n,p|m+n+p=5}\left(\nabla J\right)_{e^m \psi^n \bar{\psi}^{p}}
	\end{align}
	Imposing $\nabla J=0$ implies imposing $\left(\nabla J\right)_{e^m \psi^n \bar{\psi}^{p}}=0$ individually on each of the terms inside the sum and we will refer to this as the $e^m\psi^n\bar{\psi}^{p}$ Bianchi identities. The $e^m\psi^{p}\bar{\psi}^{n}$ Bianchi identity, for $n\neq p$ is related to the  $e^m\psi^n\bar{\psi}^{p}$ Bianchi identity by conjugation and would be referred to as the conjugate Bianchi, whereas the $e^m\psi^{n}\bar{\psi}^{n}$ Bianchi would be self-conjugate.
	\subsection{Solving the $\psi^5$ and conjugate Bianchi identities}
	One can easily notice that this Bianchi will only come from $\nabla_{1/2}J_{\psi^4}$, which can be evaluated as shown below:
	\begin{align}\label{3.15}
	\nabla_{1/2}J_{\psi^{4}}&=\frac{1}{8}\bar{\psi}^{i}\psi^{j}\bar{\psi}^{k}\psi^{l}\varepsilon_{ijm}\varepsilon_{kln}\bar{\psi}^{p}\nabla_{p}C^{mn}=0
	\end{align}
	Note that $\nabla_{p}C^{mn}$ belongs to $\bf{\bar{3}}\otimes\bf{6}=\bf{15}\oplus\bf{3}$ of SU(3). The projections to $\bf{15}$ and $\bf{3}$ are as given below:
	\begin{align}\label{3.16}
	\left(\nabla_{p}C^{mn}\right)_{\bf{15}}&=\nabla_{p}C^{mn}-\frac{1}{2}\delta^{(m}_{p}\nabla_{l}C^{n)l}\nonumber \\
	\left(\nabla_{p}C^{mn}\right)_{\bf{3}}&=\frac{1}{2}\delta^{(m}_{p}\nabla_{l}C^{n)l}
	\end{align}
	Now,
	\begin{align}\label{3.17}
	\frac{1}{8}\bar{\psi}^{i}\psi^{j}\bar{\psi}^{k}\psi^{l}\varepsilon_{ijm}\varepsilon_{kln}\bar{\psi}^{p}\left(\nabla_{p}C^{mn}\right)_{\bf{3}}=\frac{1}{16}\bar{\psi}^{i}\psi^{j}\bar{\psi}^{k}\psi^{l}\varepsilon_{ijp}\varepsilon_{kln}\bar{\psi}^{p}\nabla_{q}C^{qn}=0
	\end{align}
	This is identically zero because of the following identity:
	\begin{align}\label{3.18}
	\bar{\psi}^{i}\psi^{j}\bar{\psi}^{p}\varepsilon_{ijp}=\frac{1}{2}\bar{\psi}^{p}\psi^{j}\bar{\psi}^{i}\varepsilon_{ijp}=0
	\end{align}
	We have used Fierz identities in the above equation to go from first line to the second line. As a consequence of this, we conclude that we need to impose $\left(\nabla_{p}C^{mn}\right)_{\bf{15}}=0$ so that $\nabla_{1/2}J_{\psi^4}=0$. Hence the left supersymmetry-generator should act on $C^{ij}$ as:
	\begin{align}\label{3.19}
	\nabla_{k}C^{ij}=\frac{1}{2}\delta^{(i}_{k}\hat{\rho}^{j)}, \quad \text{where,} ~ \hat{\rho}^{j}\equiv\nabla_{l}C^{jl}
	\end{align}
	Therefore, the left supersymmetry transformation of $C^{ij}$ should take the following form:
	\begin{align}\label{3.20}
	\delta^{Q}_{L}C^{ij}=\bar{\epsilon}^{k}\nabla_{k}C^{ij}=\frac{1}{2}\bar{\epsilon}^{(i}\hat{\rho}^{j)}
	\end{align}
	To summarize, the $\psi^5$ Bianchi imposes a constraint on the allowed left-supersymmetry transformation of $C^{ij}$ which should be as given in the equation above. This defines a new component $\hat{\rho}^j$ of the abstract multiplet in terms of which we are constructing our $N=3$ density formula. The $\bar{\psi}^5$ Bianchi which is a conjugate of the $\psi^5$ Bianchi will impose a constraint on the right-supersymmetry transformation of $C_{ij}$ which would be given in terms of $\hat{\rho}_{i}$ (which is a conjugate of $\hat{\rho}^{i}$)\footnote{We are following the chiral notation where objects with indices lowered are related to the objects with indices raised by conjugation. For bosonic objects such as $C^{ij}$ this is a simple hermitian conjugation i.e $C_{ij}=(C^{ij})^*$. For fermionic objects such as $\hat{\rho}^{i}$, it means $\hat{\rho}_{i}=-C^{-1}(\bar{\hat{\rho}}_{i})^T$, where $\bar{\hat{\rho}}_{i}$ is the Dirac conjugate of $\hat{\rho}^{i}$ and $C$ is the charge conjugation matrix.} as shown below:
	\begin{align}\label{3.21}
	\delta^{Q}_{R}C_{ij}=\bar{\epsilon}_{k}\nabla^{k}C_{ij}=\frac{1}{2}\bar{\epsilon}_{(i}\hat{\rho}_{j)}
	\end{align}
	\subsection{Solving the $\psi^{4}\bar{\psi}$ and conjugate Bianchi identities}
	The $\psi^4 \bar{\psi}$ Bianchi will come from $\bar{\nabla}_{1/2}J_{\psi^{4}}$ and $t_{0}J_{e\psi^{3}}$. $\bar{\nabla}_{1/2}J_{\psi^{4}}$ can be broken up into two parts: i) $t_0$ closed and ii) non-$t_0$ closed. The non-$t_0$ closed part will give us a constraint on the right supersymmetry transformation of the composite $C^{ij}$ and the $t_0$ closed part will get cancelled by adding an appropriate $J_{e\psi^{3}}$ to the 4-form $J$. By definition:
	
	\begin{align}\label{3.22}
	\bar{\nabla}_{1/2}J_{\psi^{4}}&=\frac{1}{8}\bar{\psi}^{i}\psi^{j}\bar{\psi}^{k}\psi^{l}\varepsilon_{ijm}\varepsilon_{kln}\bar{\psi}_{p}\nabla^{p}C^{mn}
	\end{align}
	where $\nabla^{p}C^{mn}$ is in the $\bf{3}\otimes\bf{6}=\bf{10}\oplus\bf{8}$ of SU(3). The projections on to the irreducible representations are given as:
	\begin{align}\label{3.23}
	\left(\nabla^{p}C^{mn}\right)_{\bf{10}}&=\nabla^{(p}C^{mn)}\nonumber \\
	\left(\nabla^{p}C^{mn}\right)_{\bf{8}}&=\frac{2}{3}\varepsilon^{lp(m}\rho^{n)}{}_{l}, \quad \text{where,}~\rho^{i}{}_{j}\equiv\varepsilon_{klj}\nabla^{k}C^{li}
	\end{align}
	Upon explicit computation, we find that if we substitute the $\bf{8}$ projection of $\nabla^{p}C^{mn}$ in (\ref{3.22}), we get a $t_0$ closed expression as shown below:
	\begin{align}\label{3.24}
	\frac{1}{8}\bar{\psi}^{i}\psi^{j}\bar{\psi}^{k}\psi^{l}\varepsilon_{ijm}\varepsilon_{kln}\bar{\psi}_{p}\left(\nabla^{p}C^{mn}\right)_{\bf{8}}=-\frac{1}{6}\varepsilon_{kln}T^{a}\bar{\psi}^{k}\psi^{l}\bar{\psi}^{m}\gamma_{a}\rho^{n}{}_{m}
	\end{align}
	This can be cancelled by $t_0$ acting on the following $J_{e\psi^3}$ (Note that our exterior derivative act from the right):
	\begin{align}\label{3.25}
	J_{e\psi^{3}}&=-\frac{1}{6}\varepsilon_{kln}e^{a}\bar{\psi}^{k}\psi^{l}\bar{\psi}^{m}\gamma_{a}\rho^{n}{}_{m}
	\end{align}
	The $\bf{10}$ projection of $\nabla^{p}C^{mn}$ gives a non-$t_0$ closed expression and has to cancel on its own. This gives us the following constraint:
	\begin{align}\label{3.26}
	\nabla^{(i}C^{jk)}=0
	\end{align}
	In terms of supersymmetry transformation, it means that the right supersymmetry transformation of $C^{ij}$ should take the following form:
	\begin{align}\label{3.27}
	\delta^{Q}_{R}C^{ij}=\frac{2}{3}\varepsilon^{lk(i}\bar{\epsilon}_{k}\rho^{j)}{}_{l}
	\end{align}
	The conjugate Bianchi will put a constraint on the left-supersymmetry transformation of $C_{ij}$ and give us a $J_{e\bar{\psi}^3}$ as shown below\footnote{Note the difference in the alignment of the indices in $\rho^{i}{}_{j}$ and $\rho_{i}{}^{j}$. In the chiral notation we are following they are related to the conjugate of each other as $\rho_{i}{}^{j}=-C^{-1}\left(\bar{\rho}_{i}{}^{j}\right)^{T}$, where $\bar{\rho}_{i}{}^{j}$ is the Dirac conjugate of $\rho^{i}{}_{j}$ and $C$ is the charge conjugation matrix.} :
	\begin{align}\label{3.28}
	\delta_{L}^{Q} C_{ij}&=\frac{2}{3}\varepsilon_{lk(i}\bar{\epsilon}^{k}\rho_{j)}{}^{l}\nonumber \\
	J_{e\bar{\psi}^{3}}&= -\frac{1}{6}\varepsilon^{kln}e^{a}\bar{\psi}_{k}\psi_{l}\bar{\psi}_{m}\gamma_{a}\rho_{n}{}^{m}
	\end{align}
	\subsection{Solving the $\psi^{3}\bar{\psi}^{2}$ and conjugate Bianchi}
	The $\psi^{3}\bar{\psi}^{2}$ Bianchi will come from $t_{1/2}J_{\psi^4}$ and $t_0 J_{e\psi^2\bar{\psi}}$. Explicit computation shows that $t_{1/2}J_{\psi^4}$ is $t_0$ closed as shown below:
	\begin{align}\label{3.29}
	t_{1/2}J_{\psi^{4}}&=-\frac{1}{2}T^{b}\bar{\psi}^{i}\psi^{j}\bar{\psi}_{k}\gamma_{b}\Lambda_{L}\varepsilon_{ijl}C^{kl}
	\end{align}
	This is cancelled by adding the following $J_{e\psi^{2}\bar{\psi}}$ to the 4-form $J$:
	\begin{align}\label{3.30}
	J_{e\psi^{2}\bar{\psi}}=-\frac{1}{2}e^{b}\bar{\psi}^{i}\psi^{j}\bar{\psi}_{k}\gamma_{b}\sigma^{kl}\varepsilon_{ijl}
	\end{align}
	where,
	\begin{align}\label{3.31}
	\sigma^{ij}=C^{ij}\Lambda_{L}
	\end{align}
	The conjugate $\bar{\psi}^{3}{\psi}^{2}$ Bianchi, would give us the conjugate $J_{e\bar{\psi}^{2}{\psi}}$ as:
	\begin{align}\label{3.32}
	J_{e\bar{\psi}^{2}{\psi}}=-\frac{1}{2}e^{b}\bar{\psi}_{i}\psi_{j}\bar{\psi}^{k}\gamma_{b}\sigma_{kl}\varepsilon^{ijl}
	\end{align}
	where,
	\begin{align}\label{3.33}
	\sigma_{ij}=C_{ij}\Lambda_{R}
	\end{align}
	\subsection{The remaining Bianchi identities and the final density formula}
	The Bianchi identities that are remained to be solved are $e\psi^4$, $e\psi^3\bar{\psi}$, $e\psi^2\bar{\psi}^2$, $e^2\psi^3$, $e^2\psi^2\bar{\psi}$, $e^3\psi^2$, $e^3\psi\bar{\psi}$, $e^4\psi$ and their conjugates. From the calculations explicitly shown in the previous sections it is clear that $e\psi^4$, $e^2\psi^3$, $e^3\psi^2$, $e^4\psi$ and their conjugate Bianchi will only give rise to constraints as the expressions coming from these Bianchi identities will be all non-$t_0$ closed. The remaining Bianchi identities will give rise to some constraints as well as generate new terms in the 4-form $J$ in order to cancel the $t_0$ closed expressions from these Bianchi identities. The constraints that will arise from all these Bianchi identities will be satisfied if the supersymmetry algebra is closed and the initial constraint (\ref{3.19}) discussed in the previous section is satisfied. The argument is similar to what is discussed in Appendix B.2 of \cite{Butter:2019edc} and Appendix-C of \cite{Hegde:2019ioy}. The new terms in the 4-form $J$ that are generated to cancel the $t_0$ closed expressions of the Bianchi identities are:
	\begin{align}\label{3.34}
	J_{e^2 \psi^2}&=-\frac{1}{2}e^a e^b \bar{\psi}^{i}\psi^{j}H^{-}{}^{l}_{ab}\varepsilon_{ijl}-\frac{1}{2}e^a e^b \bar{\psi}^{i}\gamma_{ab}\psi^{j}K_{ij}-\frac{1}{32}e^a e^b \bar{\psi}^{i}\psi^{j}G^{+}{}^{l}_{ab}\varepsilon_{ijl}\nonumber \\
	J_{e^2\psi\bar{\psi}}&=-\frac{1}{12}e^{a}e^{b}\bar{\psi}^{l}\gamma^{c}\psi_{k}\bar{\Lambda}_{L}\gamma^{d}\rho^{k}{}_{l}\varepsilon_{abcd}+\text{h.c}\nonumber \\
	J_{e^3\psi}&= \frac{1}{3}e^a e^b e^c \bar{\psi}^{k}\gamma^{d}\mathcal{N}_{k}\varepsilon_{abcd}+\frac{1}{3}e^a e^b e^c \bar{\psi}^{k}\mathcal{M}^{d}_{k}\varepsilon_{abcd}\nonumber \\
	J_{e^4}&=\frac{1}{72}e^a e^b e^c e^d \mathcal{L}\varepsilon_{abcd}
	\end{align} 
	The conjugate terms such as $J_{e^2 \bar{\psi}^2}$ and $J_{e^3\bar{\psi}}$ are obtained by taking the conjugates of $J_{e^2 {\psi}^2}$ and $J_{e^3{\psi}}$ respectively. There are some new components that have appeared in the above expressions which are related to the components of the abstract multiplet that appeared in the previous subsections as explained. The components $H^{-}{}^{l}_{ab}$, $K_{ij}$ appearing in $J_{e^2 \psi^2}$ are given as:
	\begin{align}\label{3.35}
	H^{-}{}^{l}_{ab}&=\frac{1}{2}C^{lm}T_{ab m}-\frac{1}{16}\bar{\Lambda}_{L}\gamma_{ab}\hat{\rho}^{l},\nonumber \\
	K_{ij}&=\frac{1}{24}F_{ij}+\frac{1}{4}\bar{\Lambda}_{R}\Lambda_{R}C_{ij}
	\end{align}
	$F_{ij}$ appearing above and $G^{+}{}^{l}_{ab}$ appearing in $J_{e^2 \psi^2}$ arises in the right supersymmetry transformations of $\rho^{i}{}_{j}$ as shown below:
	\begin{align}\label{right_susy_rho}
	\delta_{Q}^{R}\rho^{i}{}_{j}&=\frac{3}{4}C^{ik}E_{j}\epsilon_{k}-\frac{1}{4}\delta^{i}_{j}C^{lk}E_{k}\epsilon_{l}+\frac{3}{8}\bar{\Lambda}_{L}\hat{\rho}^{i}\epsilon_{j}-\frac{1}{8}\delta^{i}_{j}\bar{\Lambda}_{L}\hat{\rho}^{k}\epsilon_{k}\nonumber\\
	&\quad -\frac{1}{4}\varepsilon^{ikl}F_{jk}\epsilon_{l}+\frac{3}{64}\gamma\cdot G^{i}\epsilon_{j}-\frac{1}{64}\delta^{i}_{j}\gamma\cdot G^{k}\epsilon_{k}
	\end{align}
	The components $\mathcal{N}_{k}$ and $\mathcal{M}^{a}_{k}$ appearing in $J_{e^3\psi}$ are given as:
	\begin{align}\label{MN-composites}
	{\mathcal{N}}_{k}&=-\frac{1}{32}\gamma\cdot T^{l}\Lambda_{R}C_{kl}+\frac{1}{192}\tilde{\theta}_{k}+\frac{1}{4}\theta_{k}\nonumber\\
	{\mathcal{M}}_{ak}&= -\frac{1}{64}\gamma\cdot T_{l}\gamma_{a}\rho^{l}{}_{k}+\frac{1}{48}{\Upsilon}_{ak}
	\end{align}
	where, the components $\theta_k$ appears in the right-supersymmetry transformations of $K_{ij}$, $\tilde{\theta}_{k}$ and $\Upsilon_{ak}$ appears in the left-supersymmetry transformation of $\mathcal{G}_{a}{}^{i}{}_{j}\equiv \bar{\Lambda}_{L}\gamma_{a}\rho^{i}{}_{j}-\bar{\Lambda}_{R}\gamma_{a}\rho_{j}{}^{i}$ as shown below:
	\begin{align}\label{KG-transformation}
	\delta_{Q}^{R}K_{ij}&=\bar{\epsilon}_{k}\tau^{k}_{ij}+\frac{1}{2}\bar{\epsilon}_{(i}\theta_{j)}\nonumber\\
	\delta_{Q}^{L}\mathcal{G}_{d}{}^{i}{}_{j}&=\bar{\epsilon}^{k}\alpha_{d}{}^{i}_{jk}+\frac{1}{4}\bar{\epsilon}^{k}\gamma_{d}\tilde{\alpha}^{i}_{jk}+\frac{1}{2}\varepsilon_{mjk}\varepsilon^{m}\beta_{d}^{ik}+\frac{1}{8}\varepsilon_{mjk}\varepsilon^{m}\gamma_{d}\tilde{\beta}^{ik}+\frac{3}{8}\bar{\epsilon}^{i}{\Upsilon}_{dj}+\frac{3}{32}\bar{\epsilon}^{i}\gamma_{d}\tilde{\theta}_{j}-\frac{1}{8}\delta^{i}_{j}\bar{\epsilon}^{m}{\Upsilon}_{dm}\nonumber\\
	&\;\;\; -\frac{1}{32}\delta^{i}_{j}\bar{\epsilon}^{m}\gamma_{d}\tilde{\theta}_{m}
	\end{align}
	The component $\Upsilon_{aj}$ is $\gamma$-traceless i.e. $\gamma^{a}\Upsilon_{aj}=0$. Finally the component $\mathcal{L}$ appearing in $J_{e^4}$ is given as:
	\begin{align}\label{3.39}
	\mathcal{L}&=\frac{1}{2}H^{-l}\cdot T_{l}-\frac{1}{2}H^{+}_{l}\cdot T^{l}+Y-\bar{Y}
	\end{align}
	$H^{+}_{abl}$ is the conjugate of $H^{-}{}^{l}_{ab}$ defined in (\ref{3.35}), the component $Y$ appears in the right-supersymmetry transformations of $\mathcal{N}_{i}$ as shown below and $\bar{Y}$ is its hermitian conjugate:
	\begin{align}\label{3.40}
	\delta_{Q}^{R}\mathcal{N}_{i}&=-\frac{1}{2}W^{j}{}_{i}\epsilon_{j}-\frac{1}{6}Y\epsilon_{i}+\frac{1}{8}Z^{+}_{ab}{}^{j}{}_{i}\epsilon_{j}+\frac{1}{24}\tilde{Z}^{+}_{ab}\epsilon_{i}
	\end{align}
	In the next section, we will find the composite expression for $C^{ij}$ in terms of the Weyl multiplet fields that have the required properties that it should be of Weyl weight and Chiral weight $+2$, invariant under S-supersymmetry and satisfy the constraint (\ref{3.20}). The remaining terms of the 4-form $J$ are calculated by taking successive supersymmetry transformations on this $C^{ij}$ and using the expressions given in (\ref{3.25}, \ref{3.32}, \ref{3.34}). This will give us the final invariant action for $N=3$ conformal supergravity.
	\section{Invariant actions for the Weyl multiplet}\label{sec-actions}
	
	In section-\ref{sec-density}, we constructed the density formula beginning with an ansatz for the highest gravitino term,
	\begin{align}
	J_{\psi^{4}}&=\frac{1}{4}\bar{\psi}^{i}\psi^{j}\bar{\psi}^{k}\psi^{l}\varepsilon_{ijm}\varepsilon_{kln}C^{mn},\nonumber \\
	J_{\bar{\psi}^{4}}&=\frac{1}{4}\bar{\psi}_{i}\psi_{j}\bar{\psi}_{k}\psi_{l}\varepsilon^{ijm}\varepsilon^{kln}C_{mn}.
	\end{align}
	To use the density formula derived in section-\ref{sec-density} to obtain an action for the Weyl multiplet, we need to find a suitable composite $C^{ij}$ with Weyl weight $+2$, chiral weight $+2$, is invariant under S-supersymmetry and belongs to the $\mathbf{6}$ representation of $SU(3)$. There is a combination of the Weyl multiplet fields shown below that is unique upto an over-all complex rescaling as given below.
	\begin{align}\label{cdef}
	C^{ij}=\alpha(T^i\cdot T^j+\bar{\Lambda}_R\chi^{ij}+\frac{1}{4}E^iE^j),
	\end{align} 
	where $\alpha$ is a complex number. The corresponding conjugate field $C_{ij}$ is,
	\begin{align}\label{conjugatecdef}
	C_{ij}=\alpha^*(T_i\cdot T_j+\bar{\Lambda}_L\chi_{ij}+\frac{1}{4}E_iE_j).
	\end{align}
	The composites that follow from the transformation of $J_{\psi^4}$ carry the factor $\alpha$, while the composites that follow from $J_{\bar{\psi}^4}$ carry the complex conjugate factor $\alpha^*$. Thus, we see that the modulus of $\alpha$ is redundant as it leads to an over-all rescaling of the Lagrangian. We can consider two choices to account for the argument of $\alpha$. We take $\alpha=1$ and $\alpha=i$ as the basis choices which lead to two different actions for the Weyl multiplet.
	
	This is to be expected already at the bosonic level. There are two actions for conformal gravity in four dimensions, which are called the Weyl squared action and the Pontryagin density given as below.
	\begin{align}
	\mathcal{L}_{(\text{Weyl})^2}&=C_{\mu\nu\rho\sigma}C^{\mu\nu\rho\sigma}=C_{\mu\nu\rho\sigma}{}^+C_{\mu\nu\rho\sigma}{}^++C_{\mu\nu\rho\sigma}{}^-C_{\mu\nu\rho\sigma}{}^-,\nonumber\\
	\mathcal{L}_{\text{Pontryagin}}&=\varepsilon^{\mu\nu\rho\sigma}C_{\alpha\beta\mu\nu}C_{\rho\sigma}{}^{\alpha\beta}=C_{\mu\nu\rho\sigma}{}^+C_{\mu\nu\rho\sigma}{}^+-C_{\mu\nu\rho\sigma}{}^-C_{\mu\nu\rho\sigma}{}^-,
	\end{align}
	where $C^{\mu\nu\rho\sigma}$ is the Weyl tensor and $C_{\mu\nu\rho\sigma}{}^{\pm}$ denote its self-dual and anti-self dual components. This tensor can indeed be obtained using the conformal curvature as, $C_{\mu\nu\rho\sigma}=(R(M)(e,b)_{\mu\nu\rho\sigma})\vline{}_{b_\mu=0}$, where the dependent gauge field $f_{\mu a}$ has been replaced in terms of the vielbein and dilation gauge field before setting the latter to zero. In $N=2$ conformal supergravity, the action for the Weyl multiplet is built out of the chiral density formula. Weyl multiplet can be embedded either in the chiral or the anti-chiral multiplet to give the supersymmetric completion of self-dual or anti self-dual Weyl squared actions separately whose linear combinations yield the Pontryagin density and Weyl squared action as written above for the bosonic part. However in $N=3$ conformal supergravity the situation is different. The density formula constructed in section-\ref{sec-density} is not built out of a chiral multiplet. The composite $C^{ij}$ has non-trivial chiral as well as anti-chiral $Q$-supersymmetry transformations. However when embedding the Weyl multiplet into the density formula as discussed earlier, there is a choice of real or imaginary scale factor. The two choices of $\alpha=1$ and $\alpha=i$ mentioned above precisely lead to the supersymmetric completetion of Pontryagin and Weyl squared actions respectively, in the eventuality of setting $b_\mu=0$. Although we will not set $b_\mu$ to zero, we will refer to these actions by the above names for the rest of this section.   
	
	From the density formula in section-\ref{sec-density} the full Lagrangian is given in terms of the composites as,
	\begin{align}\label{full-density}
	L&=\frac{1}{72}e^a e^b e^c e^d\mathcal{L}\varepsilon_{abcd}+ \frac{1}{3}e^a e^b e^c \bar{\psi}^{k}\gamma^{d}\mathcal{N}_{k}\varepsilon_{abcd}+\frac{1}{3}e^a e^b e^c \bar{\psi}^{k}\mathcal{M}^{d}_{k}\varepsilon_{abcd}\nn\\
	&\quad-\frac{1}{12}e^{a}e^{b}\bar{\psi}^{l}\gamma^{c}\psi_{k}\bar{\Lambda}_{L}\gamma^{d}\rho^{k}{}_{l}\varepsilon_{abcd}-\frac{1}{2}e^a e^b \bar{\psi}^{i}\psi^{j}H^{-}{}^{l}_{ab}\varepsilon_{ijl}-\frac{1}{2}e^a e^b \bar{\psi}^{i}\gamma_{ab}\psi^{j}K_{ij}\nn\\
	&\quad-\frac{1}{32}e^a e^b \bar{\psi}^{i}\psi^{j}G^{+}{}^{l}_{ab}\varepsilon_{ijl}-\frac{1}{2}e^{b}\bar{\psi}^{i}\psi^{j}\bar{\psi}_{k}\gamma_{b}C^{kl}\Lambda_L\varepsilon_{ijl}-\frac{1}{6}\varepsilon^{kln}e^{a}\bar{\psi}_{k}\psi_{l}\bar{\psi}_{m}\gamma_{a}\rho_{n}{}^{m}\nn\\
	&\quad+\frac{1}{4}\bar{\psi}^{i}\psi^{j}\bar{\psi}^{k}\psi^{l}\varepsilon_{ijm}\varepsilon_{kln}C^{mn}+\thc
	\end{align}
	{In this section, we will present the results for $\mathcal{L}$ that appears in the $e^4$ term which gives us the fully supercovariant terms in the action. We will give the complete $\mathcal{L}$ to all order in fermions for the supersymmetrization of Weyl square. For the supersymmetrization of the Pontryagin density, we will only give the fully bosonic terms appearing in $\mathcal{L}$ which can be easily seen to be a total derivative. This will serve as a consistency check of our formalism. We have computed all the composites required to obtain the actions. The required composites for Pontryagin density are presented in \ref{appendix-Pontryagin} and for Weyl squared action in \ref{appendix-weylsquare}.}
	
	Note that the composite $\mathcal{L}$ is purely imaginary by construction and $\varepsilon_{abcd}$ is purely imaginary in our convention, the $e^4$ term is purely real. {All the purely bosonic terms appearing in $i\mathcal{L}$ for the Pontryagin density can be obtained by setting $\alpha=1$ and the result is as follows.}
	\begin{align}
	i\mathcal{L} &= -24iR(M)^{abcd}R(M)^+_{abcd}-48iR(V)^+{}^j{}_i\cdot R(V)^+{}^i{}_j +144i R(A) \cdot R^+(A)
	\nn \\
	&\quad -24i T^{abi} D_a D^c T_{bci} -\frac{3i}{2} E^i D^a D_a E_i + h.c.
	\end{align}
	From the embedding with $\alpha=i$ detailed in appendix-\ref{appendix-weylsquare}, we obtain $i\mathcal{L}$ for the Weyl squared action as,
	\begin{align}
	i\mathcal{L}&= 24R(M)^{abcd}R(M)^+_{abcd}  +48R(V)^+{}^j{}_i\cdot R(V)^+{}^i{}_j -144 R(A) \cdot R^+(A)\nn \\
	&\quad -6R(V){}^j{}_i\cdot T^iE_j +12iR(A)\cdot T^iE_i +\frac{3}{16}T^i\cdot T^jE_iE_j +\frac{1}{48}D^i\;_j D^j\;_i
	\nn\\
	&\quad +\frac{3}{2} E^i D^a D_a E_i + 24 T^{abi} D_a D^c T_{bci} - \frac{1}{128}E_iE^iE_jE^j  + \frac{3}{8} \Big(T^i \cdot T^j \Big) \Big(T_i \cdot T_j \Big)
	\nonumber \\
	&\quad +48\overline{R(S)}^i\cdot R(Q)_i- \frac{3}{8}\bar{\chi}^{ij}\slashed{D} \chi_{ij}-\frac{1}{4}\bar{\zeta}_i\slashed{D}\zeta_i-12\bar{\Lambda}_R\slashed{D}D^2\Lambda_L-12\bar{\Lambda}_RD^2\slashed{D}\Lambda_L\nn\\
	&\quad + \frac{1}{8} \bar{\chi}_{ij}\zeta^i E^j  -\frac{1}{4}\bar{\zeta}_i\slashed{D}\Lambda_LE^i+\frac{1}{4}\bar{\Lambda}_L\slashed{D}E^j\zeta_j  - \frac{1}{4}\bar{\Lambda}_R \slashed{D}\zeta^i E_i  + {\frac{5}{12}} \bar{\zeta}_i \gamma \cdot T^i \slashed{D} \Lambda_L  \nn\\
	&\quad- {\frac{7}{12}}\bar{\zeta}_i \gamma \cdot \slashed{D} T^i \Lambda_L  - {\frac{1}{12} }\bar{\Lambda}_R \gamma^a \gamma \cdot T_i D_a \zeta^i - {16} \bar{\Lambda}_L \slashed{D} R(Q)_i \cdot T^i
	\nn \\
	&\quad + 2 \bar{\Lambda}_L \slashed{D} T^i \cdot R(Q)_i -14 T^i\cdot \overline{R(Q)}_i\slashed{D}\Lambda_L + 9i \bar{\Lambda}_R\gamma^a\gamma\cdot R(A)D_a\Lambda_L
	\nonumber\\
	&\quad- 48 i\bar{\Lambda}_L \gamma^a\gamma\cdot R(A)D_a\Lambda_R -39i \bar{\Lambda}_R\gamma\cdot R(A)\slashed{D}\Lambda_L - {6i} \bar{\Lambda}_R \gamma^a \Lambda_L D^b R(A)_{ab} 
	\nn \\
	&\quad +\frac{3}{16} \bar{\Lambda}_L \chi_{ij}E^i E^j +\frac{3}{4}\Lambda_L \chi_{ij} T^i \cdot T^j   + \frac{1}{16} \bar{\Lambda}_L \slashed{D} \Lambda_R E_i E^i - \frac{5}{4}\overline{D_a\Lambda}_R \gamma^a \gamma \cdot T_i  \Lambda_L E^i
	\nonumber\\
	&\quad-\frac{11}{4}\bar{\Lambda}_R\slashed{D}\gamma^{ab}\Lambda_LT_{abi}E^i +12\bar{ \Lambda}_L\gamma^cD_b\Lambda_RT_{aci}T^{abi}-12\bar{ \Lambda}_L\gamma^c\Lambda_RD_bT^{abj}T_{acj}
	\nn \\
	&\quad -\frac{5}{4}\bar{\Lambda}_R\slashed{D}\gamma\cdot T_i\Lambda_LE^i + \frac{23}{36} \bar{\Lambda}_L \slashed{D}E^i \Lambda_R E_i + \frac{1}{4}\bar{\Lambda}_R \slashed{D} E^i  \gamma \cdot T_i \Lambda_L - {\frac{1}{6}}\bar{\Lambda}_L\zeta^i\bar{\Lambda }_R\zeta_i \nn\\
	&\quad -  {  36 }\bar{\Lambda}_R \Lambda_R \bar{\Lambda}_L D^aD_a \Lambda_L  + 4 \bar{\Lambda}_R\Lambda_R \overline{D_a\Lambda}_L  \gamma^{ab} D_b\Lambda_L  + 28 \bar{\Lambda}_R\Lambda_R \overline{D_a\Lambda}_L D^a\Lambda_L \nn \\
	&\quad + {8 } \bar{ \Lambda}_RD^a\Lambda_R  \bar{\Lambda}_LD_a\Lambda_L + {40}\bar{\Lambda}_R D_a\Lambda_R\bar{ \Lambda}_L \gamma^{ab} D_b \Lambda_L  -{ 16 }\bar{ \Lambda}_R \gamma^{ab}D_b \Lambda_R \bar{ \Lambda}_L \gamma_{ac} D^c\Lambda_L \nn\\
	&\quad  - {\frac{1}{2}}\bar{\Lambda}_L\Lambda_L\bar{\Lambda}_R\zeta_jE^j
	-{\frac{1}{2}}\bar{\Lambda}_L\Lambda_L\bar{\Lambda}_R\gamma\cdot T^i\zeta_i 
	-12\bar{\Lambda}_R\Lambda_R\bar{\Lambda}_LR(Q)^i\cdot T_i  + h.c.
	\end{align}
{Apart from the supersymmetrization of the Pontryagin density being a total derivative,} the Lagrangians presented here as well as the composites appearing in \eqref{full-density} presented in the appendices obey several consistency checks coming from the Bianchi identities discussed in section-\ref{sec-density}. If the constraints from the top most layer is satisfied by the embedding then constraints from the Bianchi identities in the following layers follow from the superconformal algebra as discussed in \cite{Butter:2019edc, Hegde:2019ioy}. We have explicitly verified the constraints on the transformation of $C^{ij}$ and the transformation of $\rho^i{}_j$ that follow from five and four gravitini Bianchi identities respectively. There is an alternate check offered by the off-shell reduction of $N=4$ Weyl multiplet to $N=3$ Weyl multiplet. We will discuss this in the next section.  
	
	\section{Off-shell reduction of the Weyl multiplet from $N=4$ to $N=3$}\label{sec-reduction}
	\begin{table}
		\caption{Independent fields of the $N=4$ Weyl multiplet}\label{Table-Weyl-4}
		\begin{center}
			\begin{tabular}[t]{ | C{2cm}|C{2cm}|C{3cm}|C{2cm}|C{2cm}| }
				\hline
				Field & SU(4) Irreps & Restrictions &Weyl weight (w) & Chiral weight (c) \\ \hline
				$e_{\mu}{}^{a}$ & $\bf{1}$ & Vielbein & -1 & 0 \\ \hline
				$V_{\mu}{}^{I}{}_{J}$ & $\bf{15}$ & $(V_{\mu}{}^{I}{}_{J})^{*}\equiv V_{\mu}{}_{I}{}^{J}=-V_{\mu}{}^{J}{}_{I}$ SU(4)$_R$ gauge field &0 & 0  \\ \hline
				$b_{\mu}$ & $\bf{1}$ & dilatation gauge field &0 & 0  \\ \hline
				$\phi_\alpha$ & $\bf{1}$ & $(\phi_1)^*=\phi^1$, $(\phi_2)^*=-\phi^2$, $\phi^\alpha\phi_\alpha=1$ &0 & -1  \\ \hline
				$T^{IJ}_{ab}$ & $\bf{6}$ & Anti self-dual i.e., $T^{IJ}_{ab}=\frac{1}{2}\varepsilon_{abcd}T^{IJ}{}^{cd}$ &1 & -1  \\ \hline
				$E_{IJ}$ &$\bf{\overline{10}}$ & Complex & 1&1\\ \hline
				$D^{IJ}{}_{KL}$ & $\bf{20^\prime}$ & $D_{IJ}{}^{KL}\equiv (D^{IJ}{}_{KL})^*=D^{KL}{}_{IJ}$ &2 & 0  \\ \hline
				$\psi_{\mu}{}^{I}$ & $\bf{4}$ & $\gamma_{5}\psi_{\mu}{}^{I}=\psi_{\mu}{}^{I}$&-1/2 & -1/2  \\ \hline
				$\chi^{IJ}{}_K$ & $\bf{20}$ & $\gamma_{5}\chi^{IJ}{}_K=\chi^{IJ}{}_K$ &3/2 & -1/2  \\ \hline
				$\Lambda_{I}$ & $\bar{\bf{4}}$ & $\gamma_{5}\Lambda_{I}=\Lambda_{I}$ &1/2 &-3/2 \\ \hline
			\end{tabular}
		\end{center}
	\end{table}
	In this section, we will present an off-shell reduction of the $N=4$ Weyl multiplet to the $N=3$ Weyl multiplet. The independent fields of the $N=4$ Weyl multiplet are described in Table-\ref{Table-Weyl-4}. It possesses an $SU(4)$ $R$-symmetry appropriate for the $SU(2,2|4)$ algebra. An auxiliary $U(1)$ R-symmetry has been added to the algebra so that the scalar sector can be described by an $SU(1,1)$ valued scalar $\phi_{\alpha}$. One can gauge fix this additional $U(1)$ and obtain the description in terms of the physical complex scalar $\tau$ that parametrizes the $SU(1,1)/U(1)$ coset. The gauge field for the $SU(4)$ $R$-symmetry is $V_\mu{}^I{}_J$ where $I,J=1,\ldots,4$. The gauge field $a_{\mu}$ corresponding to the auxiliary $U(1)$-symmetry is composite and is determined in terms of the independent fields given in Table-\ref{Table-Weyl-4} by solving the constraint,
	\begin{align}\label{phiconstraint}
	\phi^\alpha D_\mu\phi_\alpha=-\frac{1}{4}\bar{\Lambda}^I\gamma_\mu\Lambda_{I},
	\end{align} 
	where $\phi^\alpha$ is related to $\phi_{\alpha}$ by complex conjugation $\phi^{\alpha}=\eta^{\alpha\beta}\phi_{\beta}^{*}$ and obeys the constraint $\phi_{\alpha}\phi^{\alpha}=1$. The three real components of $\phi_{\alpha}$ parametrizes an $SU(1,1)$ manifold and $SU(1,1)$ acts linearly on them. The full multiplet contains $128+128$ independent off-shell degrees of freedom. Transformation rule for the $N=4$ Weyl multiplet can be found in \cite{Bergshoeff:1980is,Ciceri:2015qpa,Butter:2019edc}.
	
	To reduce this multiplet to the $N=3$ Weyl multiplet, we need to set the fourth supersymmetry to zero. We begin by demanding,
	\begin{align}\label{epsilon-reduction}
	\epsilon^4=0=\psi_\mu^4,
	\end{align}
	and follow through the transformation rule of the $N=4$ Weyl multiplet to obtain the conditions on the fields. An analogous procedure was followed in \cite{Yamada:2019ttz} for the off-shell reduction from $N=2$ to $N=1$ conformal supergravity.  Similar analysis was performed on the supercurrent multiplet of $N=4$ SYM to construct the $N=3$ Weyl multiplet from the current multiplet procedure in \cite{vanMuiden:2017qsh}. Here, we investigate the reduction of the $N=4$ Weyl multiplet fields directly.
	
	We will describe a couple of steps in this procedure before presenting the results. We will employ the notation $I=\{4,i\}$ where $I$ and $i$ are $SU(4)$ and $SU(3)$ indices respectively. The transformation of the vielbein reduces in a straight forward manner to,
	\begin{align}
	\delta_{Q,S}e_\mu^a=\bar{\epsilon}^i\gamma^a\psi_{\mu i}+\thc
	\end{align}
	Transformation of the gravitino field $\psi_\mu^I$ has non-trivial information. The transformation prior to the reduction is given as,
	\begin{align}
	\delta_{Q,S}\psi_\mu^I=2\mathcal{D}_\mu\epsilon^I-\frac{1}{2}\gamma\cdot T^{IJ}\gamma_\mu\epsilon_J+\varepsilon^{IJKL}\bar{\psi}_{\mu L}\epsilon_J\Lambda_K-\gamma_\mu\eta^I,
	\end{align}
	where,
	\begin{align}
	\mathcal{D}_\mu\epsilon^I&=\partial_{\mu}\epsilon^I-\frac{1}{4}\gamma\cdot \omega_\mu\epsilon^I+\frac{1}{2}(b_\mu+ia_\mu)\epsilon^I-V_\mu{}^I{}_J\epsilon^J.
	\end{align}
	In \eqref{epsilon-reduction}, we had set $\psi_\mu^4$ to zero. Demanding consistency with the above transformation rule gives us the conditions, 
	\begin{align}\label{TVreduction}
	T_{ab}{}^{i4}=0=V_\mu{}^4{}_i\;,\; \Lambda_{i}=0\;.
	\end{align}
	Identifying $\psi_\mu^i$ as the gravitino in the $N=3$ Weyl multiplet and comparing the transformation of with \eqref{N3susy} gives,
	\begin{align}
	T_{ab}{}^{ij}&=-\frac{1}{4}\varepsilon^{ijk}T_{abk},\nn\\
	\Lambda_{4}&=\Lambda_L.
	\end{align}
	where on the RHS we have fields that belong to the $N=3$ Weyl multiplet. While deducing this, we have used,
	\begin{align}
	\varepsilon^{ijk4}:=\varepsilon^{ijk}.
	\end{align}
	Using \eqref{TVreduction} on the transformation of $V_\mu{}^4{}_i$ we can obtain the condition,
	\begin{align}
	P_a=\varepsilon_{\alpha\beta}\phi^\alpha D_a\phi^\beta=0.
	\end{align}
	This can be achieved by setting the scalars $\phi^\alpha$ to constant values consistent with the condition $\phi^\alpha\phi_\alpha=1$. We can choose either the truncation of the scalar fields as given below or any other truncation related to it by a constant $SU(1,1)$ transformation. The precise details of the scalar field truncation does not matter for the truncation of the $N=4$ Weyl multiplet to $N=3$,
	\begin{align}\label{phireduction}
	\phi_1=1, \phi_2=0.
	\end{align}
	The above condition is also consistent with $\phi_\alpha$ transformation since $\Lambda_{i}=0$ (\ref{TVreduction}). From \eqref{phireduction}, we can solve the constraint \eqref{phiconstraint}, to obtain $a_\mu$ as,
	\begin{align}
	a_\mu=\frac{i}{4}\bar{\Lambda}_R\gamma_\mu\Lambda_L.
	\end{align}
	Using the above conditions and proceeding similarly, we can infer the full reduction to the $N=3$ Weyl multiplet which can be summarized as follows. On LHS we have $N=4$ quantities and on RHS we have $N=3$ quantities:
	\begin{align}\label{5.12}
	&T_{ab}{}^{ij}= -\frac{1}{4}\varepsilon^{ijk}T_{abk}\;, T_{ab}{}^{4i}=0\;,  \psi_{\mu}^{i}= \psi_{\mu}^{i}\;, \psi_{\mu}^{4}=0\;,\nn\\ &\phi_{\mu}^{i}=\phi_{\mu}^{i}\;, \phi_{\mu}^{4}=0\;, \omega_{\mu}{}^{ab}=\omega_{\mu}{}^{ab}\;, f_{\mu}^{a}= f_{\mu}^{a}\;, b_{\mu}= b_{\mu}\;,a_\mu=\frac{i}{4}\bar{\Lambda}_R\gamma_\mu\Lambda_L\;, \nonumber \\
	&V_{\mu}{}^{4}{}_{4}= \frac{3i}{2}A_{\mu}+\frac{3}{8}\bar{\Lambda}_{R}\gamma_{\mu}\Lambda_{L}\;, V_{\mu}{}^{i}{}_{j}= V_{\mu}{}^{i}{}_{j}-\frac{i}{2}\delta^{i}{}_{j}A_{\mu}-\frac{1}{8}\delta^{i}_{j}\bar{\Lambda}_{R}\gamma_{\mu}\Lambda_{L}\;,\nn\\ 
	&\phi_1= 1\;, \phi_2=0\;, \Lambda_{4}=\Lambda_{L}\;, \Lambda_{i}=0\;, E_{4j}=-\frac{1}{4}E_{j}\;, E_{ij}=0\;,\nonumber \\
	& \chi^{4}_{4j}=\frac{1}{24}\zeta_{j}+\frac{1}{24}E_{k}\Lambda_R\;, \chi^{i}_{jk}=-\frac{1}{16}\varepsilon_{jkm}\chi^{im}-\frac{1}{24}\delta^{i}_{[j}\zeta_{k]}-\frac{1}{24}\delta^{i}_{[j}E_{k]}\Lambda_R\;, \nonumber \\
	&\chi^{4}_{ij}=0\;, \chi^{i}_{4j}=0\;, D^{4i}{}_{4j}=-\frac{1}{48}D^{i}{}_{j}\;, D^{ij}{}_{kl}=\frac{1}{12}\delta^{[i}{}_{[k}D^{j]}{}_{l]}\;, D^{4i}{}_{jk}=0\;, D^{ij}{}_{4k}=0.
	\end{align}
	Upon using the above reduction, transformation rule for the $N=3$ Weyl multiplet is reproduced. This while interesting in its own right, also provides us with a check for our results from section-\ref{sec-actions}. The most general action for $N=4$ Weyl multiplet was constructed recently in \cite{Ciceri:2015qpa,Butter:2016mtk,Butter:2019edc}. The action is characterized in terms of an arbitrary holomorphic function $\mathcal{H}(\phi_\alpha)$. The action also contains $SU(1,1)$ derivatives of the holomorphic function. The part of the action that comes with $\mathcal{H}$ has zero $U(1)$ weight whereas the parts that come with the $SU(1,1)$ derivatives of $\mathcal{H}$ has non trivial $U(1)$-weight in order to cancel the $U(1)$-weights of the $SU(1,1)$ derivatives of $\mathcal{H}$. Under the truncation described above, the holomorphic function along with all its derivatives is set to a constant. Although the derivatives of the function $\mathcal{H}$ is set to a constant under the truncation, the terms that accompany it in the $N=4$ Lagrangian vanish. This is because the terms that accompany the derivatives of $\mathcal{H}$ have a non-trivial $U(1) $-weight while they are invariant under the $SU(4)$-R symmetry. And such terms should necessarily vanish under the truncation. This can be argued as follows. One can easily check that under the above mentioned truncation of the $N=4$ Weyl multiplet to the $N=3$ Weyl multiplet, the $SU(4)$ and the $U(1)$ parameters of the $N=4$ soft-supersymmetry algebra \cite{Bergshoeff:1980is,Butter:2019edc} are related to the $SU(3)$ and $U(1)$ parameters of the $N=3$ soft-supersymmetry algebra \cite{vanMuiden:2017qsh,Hegde:2018mxv} as follows\footnote{The $N=3$ soft supersymmetry algebra in our convention differs from that of \cite{vanMuiden:2017qsh,Hegde:2018mxv} and is given in (\ref{A7}, \ref{A8}).}(On the LHS we have $N=3$ parameters and on the R.H.S we have $N=4$ parameters):
	\begin{align}\label{5.13}
	\lambda^{i}{}_{j}&=\Lambda^{i}{}_{j}+\frac{1}{3}\delta^{i}{}_{j}\Lambda^{4}{}_{4}\nonumber\\
	\lambda_T&=\Lambda_T-\frac{2i}{3}\Lambda^{4}{}_{4}
	\end{align}
	Hence, in order to be consistent with the soft-supersymmetry algebra, any field or combination of fields in $N=4$ which survives the truncation to $N=3$ should have such an $SU(4)$ and $U(1)$ transformation in $N=4$ conformal supergravity, which upon truncation to $N=3$ should yield:
	\begin{align}\label{5.14}
	\delta_{SU(4)}(\Lambda)+\delta_{U(1)}(\Lambda_T)=\delta_{SU(3)}(\lambda)+\delta_{U(1)}(\lambda_T)\;,
	\end{align}
	where $\lambda$ and $\lambda_T$ are related to $\Lambda$ and $\Lambda_T$ exactly as given in (\ref{5.13}). For example, this indeed holds for $T_{abl}=-2\varepsilon_{ijl}T_{ab}{}^{ij}$ as shown in the calculations below:
	\begin{align}
	&\left(\delta_{SU(4)}(\Lambda)+\delta_{U(1)}(\Lambda_T)\right)T_{ab}^{ij}=-2\Lambda^{[i}{}_{k}T_{ab}^{j]k}-i\Lambda_TT_{ab}^{ij}\nonumber \\
	&\implies \left(\delta_{SU(4)}(\Lambda)+\delta_{U(1)}(\Lambda_T)\right)T_{abl}=-2\varepsilon_{ijl}\left(\delta_{SU(4)}(\Lambda)+\delta_{U(1)}(\Lambda_T)\right)T_{ab}^{ij}\nonumber \\
	&=4\varepsilon_{ijl}\Lambda^{i}{}_{k}T_{ab}^{jk}-i\Lambda_T T_{abl}=\Lambda^{i}{}_{i}T_{abl}-\Lambda^{i}{}_{l}T_{abi}-i\Lambda_T T_{abl}\nonumber \\
	&=-\lambda^{i}{}_{l}T_{abi}-i\lambda_TT_{abl}=\left(\delta_{SU(3)}(\lambda)+\delta_{U(1)}(\lambda_T)\right)T_{abl}
	\end{align}
	In the last line we have used $\Lambda^{i}{}_{i}=-\Lambda^{4}{}_{4}$ and (\ref{5.13}). However if we have a combination of fields which is invariant under $SU(4)$ and has a non trivial $U(1)$-weight, then (\ref{5.14}) will never be satisfied on it and hence it wont survive the truncation. For example $\varepsilon^{IJKL}T_{IJ}\cdot T_{KL}$ does not transform under $SU(4)$ but has a non trivial $U(1)$-weight and one can easily check that it does not survive the truncation (\ref{5.12}).
	Thus we see that the terms that accompany the $SU(1,1)$ derivatives of $\mathcal{H}$ must vanish upon using the truncation (\ref{5.12}). Terms that accompany $\mathcal{H}$ survive and as $\mathcal{H}$ reduces to a constant, one can choose it to be $1$ or $i$. {Upon using $\mathcal{H}=i$ one would get the supersymmetrization of Pontryagin density in $N=4$ which, as expected, gives a total derivative. Whereas upon using $\mathcal{H}=1$ and  the truncations in (\ref{5.12}), gives us the Weyl squared action discussed in section-\ref{sec-actions} up to a total derivative and an overall multiplicative factor. }This serves as a non trivial check on our results.
	
	\section{Conclusion and future directions}\label{sec-conclusion}
	Superconformal tensor calculus provides a systematic method to construct general matter coupled Poincar\'e supergravity theories upto six spacetime dimensions.. In four dimensions, while the invariants for $N=1,2,4$ supergravity are relatively well studied, a superconformal approach to study $N=3$ supergravity has been lacking. In \cite{vanMuiden:2017qsh, Hegde:2018mxv}, the Weyl multiplet in $N=3$ conformal supergravity was constructed.

	In this paper, we have constructed invariant actions for the Weyl multiplet in $N=3$ conformal supergravity in four dimensions. In the eventuality of Poincare gauge fixing which involves setting $b_\mu=0$, these actions become the Pontryagin density and the Weyl squared action. To carry out a systematic investigation of $N=3$ matter coupled actions  we need to construct the matter multiplets in $N=3$ conformal supergravity. This could be done from an off-shell reduction of the kind described in section-\ref{sec-reduction}. The off-shell reduction could also be further investigated to obtain a reduction of $N=3$ to $N=2$ conformal supergravity. Along with the results of \cite{Yamada:2019ttz}, this will connect all the conformal supergravity theories via off-shell reductions.  This is a work in progress. 
	
	Construction of new higher derivative invariant for $N=3$ supergravity is interesting also from the perspective of AdS/CFT. As mentioned earlier, $N=3$ supergravity has already been used in the context of AdS/CFT in \cite{Karndumri:2016miq, Karndumri:2016tpf}. Recently, for the case of $N=2$ supergravity, it was found in \cite{Bobev:2020egg} that higher derivative supergravity is efficient to determine subleading corrections to the large $N$ behavior of supersymmetric partition functions via the AdS/CFT correspondence. Further study on higher derivative invariants in $N=3$ supergravity may shed light on similar features for theories with $12$ supercharges.
	
	There has been some interesting works on conformal supergravity with $N=0,1,2,4$ supersymmetries in four dimensions using the double copy prescription \cite{Johansson:2017srf,Johansson:2018ues}. As per the discussion in \cite{Johansson:2017srf,Johansson:2018ues}, the minimal $N=0,1,2,4$ conformal supergravity in four dimensions has the following double copy structure
	\begin{align}
	(N=0,1,2,4)\; \text{minimal conformal supergravity}=\left(DF\right)^2_{min}\otimes (N=0,1,2,4 SYM)\;,
	\end{align}
	where $(DF)^2$ is a gauge theory that is built purely out of dimension-6 operators (counting as per $6$ space-time dimensions) and $(DF)^2_{min}$ contains solely the kinetic terms of $(DF)^2$. It would be interesting to find out if $N=3$ conformal supergravity as discussed in this paper can be obtained via a double copy prescription of the form above where we replace the second copy by a $N=3$ Super Yang Mills. $N=3$ Super yang Mills is interesting in its own right. CPT invariance implies that any field theory with $N=3$ supersymmetry necessarily has $N=4$ supersymmetry. Hence there has been very little study on field theories with $N=3$ supersymmetry in four dimensions and in order to study such theories people have resorted to either non-Lagrangian philosophy \cite{Aharony:2015oyb} or used harmonic superspace \cite{Zupnik:2000ip,Devchand:1993rt}. This poses real challenge for studying the double copy formulation of $N=3$ conformal supergravity. Another interesting aspect from the study of \cite{Johansson:2017srf,Johansson:2018ues} is that the non minimal $N=4$ conformal supergravity of Berkovits-Witten type \cite{Berkovits:2004jj} is obtained using the above double copy prescription where the minimal $(DF)^2$ theory is replaced by the non-minimal one and this coincides with the $N=4$ conformal supergravity action of \cite{Butter:2016mtk,Butter:2019edc} with a specific choice of the holomorphic function $\mathcal{H}(\phi_{\alpha})$. For lower supersymmetries if we replace the minimal $(DF)^2$ by a non minimal one, one obtains versions of $N=0,1,2$ conformal supergravity which arises from supersymmetry truncation of Berkovits-Witten type $N=4$ conformal supergravity. These are basically $N=0,1,2$ conformal supergravities coupled to additional matter multiplets such as vector multiplet or hypermultiplet and typically the Weyl square term would be multiplied by an axion-dilaton coming from these matter multiplets. It would be interesting to see what is the supersymmetry truncation of Berkovits-Witten type $N=4$ conformal supergravity to $N=3$. We leave to address these issues for the future.
	\section*{Acknowledgements}
      This work is partially supported by SERB core research grant grant CRG/2018/002373, Government of India. 
	
	\appendix
	\section{Conventions}\label{appendix-conventions}
	In this section, we will explain the conventions used for the superconformal transformations and the soft algebra. The $N=3$ superconformal algebra has an $SU(3)\times U(1)$ $R$-symmetry. The $SU(3)$ indices are denoted by indices $i,j,k,\ldots$ throughout the paper.  The $U(1)$ $R$-charge is called the chiral weight and let us denote it as $c$. Consider a field $A^i$ which transforms in the fundamental representation of $SU(3)$. Its $SU(3)$ transformation is given by, 
	\begin{align}
	\delta_V A^i=\lambda^i{}_jA^j.
	\end{align}
	Therefore, its covariant derivative is given by,
	\begin{align}
	D_\mu A^i=-V_\mu{}^i{}_jA^j+\ldots,
	\end{align}
	where the dots indicate covariantisation with respect to the other superconformal transformations. Note that, in \cite{Hegde:2018mxv}, the gauge field was used with the first index down, as $V_{\mu j}{}^i$. The two conventions are related by the equation, $V_{\mu j}{}^i=-V_\mu{}^i{}_j$.
	
	For the $U(1)$ $R$-symmetry, our convention is such that the left-chiral gravitino $\psi_{\mu}^{i}$, has chiral weight $c=-1/2$. That is, we have reversed all the chiral weights relative to that of \cite{Hegde:2018mxv}. This is achieved by replacing $A_\mu$ by $-A_\mu$ and the $U(1)$ transformation parameter $\lambda_T$ by $-\lambda_T$ wherever they appear. 
	
	For the local Lorentz transformation, our convention is as follows. Consider a field $B^a$ where $a$ is the local Lorentz index. Its transformation under local Lorentz transformation is given as,
	\begin{align}
	\delta_M B^a=\lambda_M{}^a{}_bV^b,
	\end{align}
	which has a minus sign relative to that of \cite{Hegde:2018mxv}. This is achieved by replacing $\omega_\mu^{ab}$ by $-\omega_{\mu}^{ab}$ and the local Lorentz trnasformation parameter $\lambda_M^{ab}$ by $-\lambda_M^{ab}$, wherever they appear. Consequently, a Majorana fermion $\chi$ has the local Lorentz transformation,
	\begin{align}
	\delta_M\chi=\frac{1}{4}\gamma\cdot \lambda_M\chi.
	\end{align}
	Further, we would like to use a convention that is standard in $N=2$ and $N=4$ conformal supergravity, where the gauge fields $\psi_{\mu}^{i}$ and $\phi_{\mu}^{i}$ transforms to $2\mathcal{D}_{\mu}\epsilon^{i}$ and $2\mathcal{D}_{\mu}\eta^{i}$ under Q and S supersymmetry respectively. This is different compared to the convention followed in \cite{Hegde:2018mxv}. In order to achieve this, we do the following redefinitions on the convention followed in \cite{Hegde:2018mxv},
	\begin{align}
	\psi_\mu^i&\rightarrow\frac{\psi_\mu^i}{2},\nn\\
	e_\mu^a&\rightarrow\frac{1}{4}e_\mu^a,\nn\\
	\eta^i&\rightarrow 2\eta^i,
	\end{align}
	where the LHS is from \cite{Hegde:2018mxv} and RHS is for the conventions followed in this paper. The second line in the above equation is a field redefinition made for convenience. Above change in convention induces the following redefinition in the K-gauge field and the corresponding parameter (where again the LHS is from \cite{Hegde:2018mxv} and RHS is for the conventions followed in this paper):
	\begin{align}
	f_{\mu}{}^{a}\to 4f_{\mu}{}^{a}\;,\; \lambda_{K}^{a}\to 4\lambda_{K}^{a}
	\end{align}
	In the conventions detailed above, the soft algebra realized on the fields is given by,
	\begin{align}\label{A7}
	[\delta^{Q}(\epsilon_1),\delta^Q (\epsilon_2)]&=\delta^{cgct}(\xi^{\mu})+\delta^{M}(\epsilon_{1}^{ab})+\delta^{Q}(\epsilon_{3}^{i})+\delta^{S}(\eta_{1}^{i})+\delta_{SU(3)}(\lambda_{1}{}_{j}{}^{i})+\delta_{U(1)}(\lambda_{1T})+\delta_K(\lambda^a_{1K})\nonumber \\
	[\delta^{Q}(\epsilon),\delta^S (\eta)]&= \delta_{D}(\lambda_D)+\delta^{M}(\epsilon_{2}^{ab})+\delta^{S}(\eta_{2}^{i})+\delta_{SU(3)}(\lambda_{2}{}_{j}{}^{i})+\delta_{U(1)}(\lambda_{2T})+\delta_K(\lambda^a_{2K})\nonumber\\
	[\delta^{S}(\eta_1),\delta^S (\eta_2)]&=\delta_K(\bar{\eta}_2^i\gamma^a\eta_{1i}+\thc)
	\end{align}
	The field dependent transformation parameters appearing on the RHS are given as:
	\begin{align}\label{A8}
	\xi^{\mu}&=2\bar{\epsilon}_{2i}\gamma^{\mu}\epsilon_{1}^{i}+\thc\nonumber \\
	\epsilon_{1}^{ab}&=\varepsilon_{ijk}\bar{\epsilon}_{2}^{i}\epsilon_{1}^{j}T_{ab}^{k}+\thc\nonumber \\
	\epsilon_{3}^{i}&=-\varepsilon^{ijk}\bar{\epsilon}_{2j}\epsilon_{1k}\Lambda_L\nonumber \\
	\eta_{1}^{i}&=-\frac{1}{12}\bar{\epsilon}_2^{[i}\epsilon_1^{k]}\zeta_{k}+\frac{1}{32}\left(\bar{\epsilon}_{2}^{i}\gamma_{a}\epsilon_{1j}-\delta^{i}_{j}\bar{\epsilon}_{2}^{k}\gamma_{a}\epsilon_{1k}+\text{h.c}\right)\gamma^{a}\Lambda_{L}E^{j}\nonumber\\
	&\quad+\frac{1}{96}\left(\bar{\epsilon}_{2}^{i}\gamma_{a}\epsilon_{1j}-\delta^{i}_{j}\bar{\epsilon}_{2}^{k}\gamma_{a}\epsilon_{1k}+\text{h.c}\right)\gamma^{a}\zeta^j-\frac{1}{4}\bar{\epsilon}_2^{[i}\epsilon_1^{j]}E_j\Lambda_R\nonumber\\
	&\quad-\frac{1}{32}\varepsilon^{ijk}(\bar{\epsilon}_2^l\gamma_a\epsilon_{1k}+\thc)\gamma^a\chi_{jl}-\frac{1}{8}\varepsilon_{jkl}\bar{\epsilon}_2^j\epsilon_1^k\chi^{il}-\frac{1}{2}\varepsilon^{ijk}\bar{\epsilon}_{2j}\epsilon_{1k}\slashed{D}\Lambda_L\nonumber\\
	&\quad-\frac{1}{16}(\bar{\epsilon}_2^i\gamma_a\epsilon_{1j}-\delta^i_j \bar{\epsilon}_2^k\gamma_a\epsilon_{1k}+\thc)\gamma\cdot T^j\gamma^a\Lambda_L\nonumber\\ 
	\lambda_{1T}&=\frac{i}{6}\varepsilon_{ijk}\bar{\epsilon}_2^j\epsilon_{1}^kE^i-\frac{2i}{3}(\bar{\epsilon}_2^j\gamma^a\epsilon_{1j})\bar{\Lambda}_L\gamma_a\Lambda_R+\thc\nonumber\\
	\lambda_1{}^i{}_j &=\frac{1}{4}\varepsilon_{jpq}\bar{\epsilon}_2^p\epsilon_1^q E^i+\frac{1}{2}(\bar{\epsilon}_2^i\gamma^a\epsilon_{1j})\bar{\Lambda}_L\gamma_a\Lambda_R-\thc-\text{trace}\nonumber\\
	\lambda^a_{1K}&=-\frac{i}{3}\bar{\epsilon}_{2}^{k}\gamma_{b}\epsilon_{1k}\tilde{R}^{ab}(A)+\frac{2}{3}\bar{\epsilon}_{2}^{i}\gamma_{b}\epsilon_{1j}\tilde{R}^{ab}(V){}^j{}_i+\frac{1}{32}\bar{\epsilon}_2^{[i}\gamma\cdot T_i\gamma^{a}\gamma\cdot T^{j]}\epsilon_{1j}\nonumber \\
	&\quad -\frac{2}{3}\varepsilon^{ijk}\bar{\epsilon}_{2i}\epsilon_{1j}D_b T^{ab}_k+\thc\nonumber\\
	\lambda_D &= -\bar{\eta}_i\epsilon^i+\thc \nonumber \\
	\epsilon_2^{ab} &=-\bar{\eta}_i\gamma^{ab}\epsilon^i+\thc\nonumber\\
	\eta_{2i}&=\frac{1}{4}\varepsilon_{ijk}\bar{\epsilon}^j\gamma_a\eta^k\gamma^a\Lambda_R\nonumber\\
	\lambda_2{}^i{}_j &=-2\bar{\epsilon}^i\eta_j-\thc-\text{trace}\nonumber\\
	\lambda_{2T} &=-\frac{i}{3}\bar{\epsilon}^i\eta_i+\thc\nonumber\\
	\lambda_{2K}^a&=-\frac{1}{24}\varepsilon_{ijk}\bar{\epsilon}^i\gamma^a\gamma\cdot T^j\eta^k+\thc
	\end{align}
		
	\section{Composites for Pontryagin density}\label{appendix-Pontryagin}
	In this section, we will present composites for the case of Pontryagin density. As discussed in section-\ref{sec-actions}, the beginning composite for the Pontryagin density is given as,
	\begin{align}
	C^{ij}=T^i\cdot T^j+\bar{\Lambda}_R\chi^{ij}+\frac{1}{4}E^iE^j.
	\end{align}
	This composite should satisfy the constraint on their supersymmetry transformation obtained from section-\ref{sec-density}. Indeed, one finds that their supersymmetry transformation takes the following form
	\begin{align}\label{csusy}
	\delta C^{ij}=\frac{2}{3}\varepsilon^{lk(i}\bar{\epsilon}_{k}\rho^{j)}{}_{l}+\frac{1}{2}\bar{\epsilon}^{(i}\hat{\rho}^{j)},
	\end{align}
	where,
	\begin{align}\label{rhodef}
	\hat{\rho}^i&= - 4\slashed{D} \gamma \cdot T^i \Lambda_R - 4T^{abi} \slashed{D} \gamma_{ab} \Lambda_R
	- 4 E^i \slashed{D} \Lambda_R -4 \slashed{D} E^i \Lambda_R  \nonumber\\
	&\quad - \frac{4}{3} \bar{\Lambda}_R\Lambda_R \zeta^i   - 8 \bar{\Lambda}_R \Lambda_R E^i \Lambda_L, \\
	\rho^i{}_l&= \frac{3}{8}\varepsilon_{jkl}\gamma\cdot T^k\chi^{ij}-\frac{1}{8}(\gamma\cdot T^i\zeta_l-\frac{1}{3}\delta^i_l\gamma\cdot T^k\zeta_k)+ 12(T^{abi}R(Q)_{abl}-\frac{1}{3}\delta^i_lT^{abk}R(Q)_{abk})\nonumber\\
	&\quad+ \frac{3}{8}\varepsilon_{jkl}E^k\chi^{ij}+\frac{3}{8}(E^i\zeta_l-\frac{1}{3}\delta^i_lE^k\zeta_k)  + 12\gamma\cdot R(V){}^i{}_l\Lambda_R-\frac{1}{2}D^l_i\Lambda_R\nonumber\\
	&\quad-\frac{3}{4}(\gamma\cdot T^i\Lambda_RE_l-\frac{1}{3}\delta^i_l\gamma\cdot T^kE_k\Lambda_R).
	\end{align}
	We can write the composite $H^{-l}{}_{ab}$ that appears in \eqref{full-density} as,
	\begin{align}
	H^{-l}{}_{ab}=\frac{1}{2}C^{lm}T_{abm}-\frac{1}{16}\bar{\Lambda}_L\gamma_{ab}\hat{\rho}^l.
	\end{align}   
	The composite $F_{ij}$ and $G_{ab}{}^i$ are obtained from the right-supersymmetry transformation of $\rho^i{}_j$ \eqref{right_susy_rho} as,
	\begin{align}
	F_{ij}&= 96\bar{R}(Q)^{ab}_{i}R(Q)_{abj}-\frac{3}{4}\varepsilon_{ikl}\varepsilon_{jmn}\bar{\chi}^{km}\chi^{ln}-12\bar{\Lambda}_{R}\slashed{D}\chi_{ij}+\frac{1}{2}\bar{\zeta}_{i}\zeta_{j}  +2\bar{\Lambda}_{R}\zeta_{(i}E_{j)}\nn \\
	&\quad -96\varepsilon_{kl(i}R(V){}^k{}_{j)}\cdot T^l -\varepsilon_{kl(i} E^k D^l{}_{j)} + {12} {T_i\cdot T_j \bar{\Lambda }_R\Lambda_R} + 3 {E_iE_j\bar{\Lambda }_R\Lambda_R},\\
	G_{ab}^{i}&=32\bar{R}(Q)_{abk}\chi^{ik}-\frac{2}{3}\bar{\zeta}_{k}\gamma_{ab}\chi^{ik}-\frac{32}{3}\varepsilon^{ikl}\bar{\zeta}_{k}R(Q)_{abl}-\frac{2}{9}\varepsilon^{ikl}\bar{\zeta}_{l}\gamma_{ab}\zeta_{k}-\frac{16}{3}\bar{\Lambda}_{R}\gamma_{ab}\slashed{D}\zeta^{i}\nonumber \\
	&\quad +64\varepsilon^{ijk}\bar{R}(Q)^{cd}_{j}\gamma_{ab}R(Q)_{cdk}+256\bar{\Lambda}_{R}R(S)_{ab}^{+i} + 32\varepsilon^{ijk}\bar{\Lambda}_{R}R(Q)_{abj}E_{k}\nn\\
	&\quad  - 24 E^i \bar{\Lambda}_L \slashed{D} \gamma_{ab}\Lambda_R  - 32 E^i \bar{\Lambda}_R   \gamma_{ab} \slashed{D} \Lambda_L - 8 \bar{\Lambda}_L \slashed{D} \gamma_{ab} E^i \Lambda_R +64\bar{\Lambda}_L\slashed{D}T_{ab}^i\Lambda_R\nn \\
	&\quad +32\bar{\Lambda}_L\gamma^d\gamma_{ab}\Lambda_RD^cT_{cd}^i  -\frac{128}{3}T_{ab}^i\bar{\Lambda }_L\slashed{D}\Lambda_R-\frac{256}{3}T_{ab}^i\bar{\Lambda}_R\slashed{D}\Lambda_L 
	-8\bar{\Lambda}_L\gamma^c\gamma\cdot T^i \gamma_{ab}D_c\Lambda_R
	\nn \\
	&\quad- 64 R(V)_{ab}{}^i{}_j E^j + 128 E^i R(A)_{ab} + 256 T_{c[a}{}^nR(V)^c{}_{b]}{}^i{}_n+256iT_{c[a}^iR(A)^c{}_{b]} \nn \\
	&\quad+128T^{cdi}R(M)_{cdab}-\frac{8}{3}T_{ab}^jD_j^i  + \frac{8}{3}T_{ab}{}^i E^j E_j   +  16 T_{c[a}{}^i T^c{}_{b]}{}^m E_m .
	\end{align}
	While $G_{ab}{}^i$ appears in the density \eqref{full-density}, $F_{ij}$ contributes to the composite $H_{ij}$ that appears in the density as given below.
	\begin{align}\label{H-composite}
	K_{ij}&=\frac{1}{24}F_{ij}+\frac{1}{4}\bar{\Lambda}_R\Lambda_RC_{ij},
	\end{align} 
	From the transformation \eqref{KG-transformation} we can read off the composite $\theta_i$ upto linear order in fermion
	\begin{align}
	\theta_i&=-8\gamma^{cd}R(Q)^{ab}{}_iR(M)_{abcd} +{32}R(V){}^j_i\cdot R(Q)_j+16iR(A)\cdot R(Q)_i - \varepsilon_{ijk}\gamma\cdot R(V){}^j{}_l\chi^{kl}\nn\\
	&\quad -\frac{4i}{3}\gamma\cdot R(A) \zeta_i -\frac{1}{3}\gamma\cdot  R(V){}^j_i\zeta_j + 8\varepsilon_{ijk}R(S)^j\cdot T^k-\frac{1}{24}\varepsilon_{ikl}\chi^{jk}D^l_j -\frac{1}{24}D_i^j\zeta_j
	\nonumber\\
	&\quad+\frac{1}{4} \slashed{D} \chi_{ij} E^j  +  \frac{1}{4} \varepsilon_{ijk} E^j \slashed{D} \zeta^{k}  +{\frac{1}{4}\gamma\cdot T^j\slashed{D}\chi_{ij}}+{\frac{1}{12}}\varepsilon_{ijk}\gamma\cdot T^k\slashed{D}\zeta^j- 2 D^a D_a E_i \Lambda_R\nn\\
	&\quad+8D_aD^cT_{cbi}\gamma^{ab}\Lambda_R +\varepsilon_{ijk}\gamma^{ab}T_{ca}{}^kT^c{}_b{}^j\slashed{D}\Lambda_L +\frac{3}{128}\varepsilon_{ijk}E_l\gamma\cdot T^j \chi^{kl} {- \frac{5}{128}} \varepsilon_{ijk} E_l E^j \chi^{kl}	\nn\\
	&\quad- \frac{5}{96}  E_j D^j\;_i  \Lambda_R -\frac{1}{384}\gamma\cdot T^j\zeta_jE_i-\frac{13}{384}\gamma\cdot T^jE_j\zeta_i {-\frac{1}{384}}\zeta_i E_j E^j -  {\frac{5}{384}} E_i E^j \zeta_{j}
	\nn\\
	&\quad +4\varepsilon_{ijk}\gamma^cD_aT_{bc}{}^jT^{abk}\Lambda_L+  \frac{1}{4} \varepsilon_{ijk} E^j \slashed{D} E^{k} \Lambda_L +\frac{1}{4}\varepsilon_{ijk}\gamma\cdot\slashed{D}T^k\Lambda_LE^j
	\nn\\
	&\quad+\frac{1}{4}\varepsilon_{ijk}\gamma\cdot T^k\slashed{D}E^j\Lambda_L -4R(V){}^j_i\cdot T_j\Lambda_R+8iR(A)\cdot T_i\Lambda_R +\frac{1}{4} \gamma \cdot R(V){}^j{}_i E_j \Lambda_R  \nn\\
	&\quad -2i  \gamma \cdot R(A)E_i \Lambda_R+\frac{1}{4} R(Q)_i\cdot T^jE_j-\frac{3}{4}E_iR(Q)_j\cdot T^j-\frac{5}{8}T_i\cdot T_j E^j\Lambda_R \nn\\
	&\quad-\frac{5}{8}T_i\cdot T_j \gamma\cdot T^j\Lambda_R-\frac{1}{8}E_iE_k\gamma\cdot T^k\Lambda_R - \frac{7}{96} E_i E^j E_j \Lambda_R 
	\end{align}
	The composite $\mathcal{G}_{d}{}^{i}{}_{j}$ appearing in $J_{e^2 \psi \bar{\psi}}$ is given as,
	\begin{align}
	\mathcal{G}_{d}{}^{i}{}_{j}&=\bar{\Lambda}_{L}\gamma_{d}\rho^{i}{}_{j}-\bar{\Lambda}_{R}\gamma_{d}\rho_{j}{}^{i}
	\end{align}
	As $\mathcal{G}_a{}^i{}_j$ is made of composites encountered earlier, we can derive the following composites that appear in its transformation \eqref{KG-transformation} as,
	\begin{align}\label{tildezetas}
	{\Upsilon}_{ai}&=\frac{1}{4}\gamma_a\slashed{E}_i\Lambda_L-E_{ai}\Lambda_L-\frac{1}{4}\gamma\cdot T_j\gamma_{a}\rho^j{}_i-\frac{1}{8}\gamma\cdot G_i\gamma_{a}\Lambda_R,\nonumber\\
	\tilde{\theta}_i&=\slashed{E}_i\Lambda_L-\rho^j{}_iE_j+2C_{ij}E^j\Lambda_R-2\hat{\rho}_i\bar{\Lambda}_R\Lambda_R,
	\end{align}
	where,
	\begin{align}\label{Eadefinition}
	E_{ai}&=\bar{Q}_{j}\gamma_{a}\rho^{j}{}_{i},
	\end{align}
	{and the purely bosonic part of $E_{ai}$} is given as,
	\begin{align}
	E_{ai}&=-64\varepsilon_{ijk}T_{ab}{}^jD^cT^b{}_c{}^k+16\varepsilon_{ijk}T^k\cdot D_aT^j  - 16 \varepsilon_{ijk} E^j{D^b} T_{ab}{}^k  -16 \varepsilon_{ijk}  T^{abk} D_b E^j 
	\end{align}
	Using the above and \eqref{MN-composites}, we can obtain the composites $\mathcal{M}_{ai}$ and $\mathcal{N}_i$ that appear in the density \eqref{full-density}. We obtain the composite $\tilde{\theta}_i$ {upto linear order in fermion} to be,
	\begin{align} 
	\tilde{\theta}_i&= - \frac{3}{8}\varepsilon_{ikl}  E_j \gamma \cdot T^l  \chi^{jk} - \frac{3}{8} \varepsilon_{ikl} E^l  E_j \chi^{jk} + \frac{1}{2} E_j D^j\;_i \Lambda_R - \frac{1}{24} \gamma \cdot T^j \zeta_jE_i  + \frac{1}{8} \gamma \cdot T^j E_j \zeta_i  	\nn\\
	&\quad-\frac{3}{8}\zeta_i E^j E_j + \frac{1}{8} E_i E^j \zeta_j -{16 \varepsilon_{ijk} E^j {D}^b T_{ab}{}^k \gamma^a \Lambda_L} -64\varepsilon_{ijk}\gamma^aT_{ab}{}^jD^cT^b{}_c{}^k\Lambda_L	\nn\\
	&\quad +16\varepsilon_{ijk}T^k\cdot\slashed{D}T^j\Lambda_L  -16 \varepsilon_{ijk} \gamma_a T^{abk} D_b E^j \Lambda_L  - 12 \gamma \cdot R(V){}^j{}_i E_j \Lambda_R 	\nn \\
	&\quad  -12  R(Q)_{i}\cdot T^{ j}  E_j + 4 E_i R(Q)_{ j} \cdot T^{ j}  + \frac{1}{2}  E_i E_j \gamma \cdot T^j\Lambda_R  + \frac{1}{2}  E_i E_j E^j\Lambda_R \nn \\
	&\quad+ 2 T_i \cdot T_j E^j \Lambda_R  
	\end{align}
	The composite $\mathcal{N}_i$ and its the right-supersymmetry transformation is given as follows ,
	\begin{align}
	{\mathcal{N}}_{i}&=-\frac{1}{32}\gamma\cdot T^{j}\Lambda_{R}C_{ij}+\frac{1}{192}\tilde{\theta}_{i}+\frac{1}{4}\theta_{i}\nonumber \\
	\delta_{Q}^{R}\mathcal{N}_{i}&=-\frac{1}{2}Y^{j}{}_{i}\epsilon_{j}-\frac{1}{6}Y\epsilon_{i}+\frac{1}{8}\gamma \cdot Y^{+}{}^{j}{}_{i}\epsilon_{j}+\frac{1}{24}\gamma \cdot Y^{+}\epsilon_{i}.
	\end{align}
	Using the results obtained for the RHS, $\mathcal{N}_i$ is given in terms of Weyl multiplet fields as,
	\begin{align}
	\mathcal{N}_i &=  2\gamma^{cd} R(Q)^{ab}_iR(M)_{abcd} +8 R(V){}^j{}_i \cdot R(Q)_j - 4iR(A)\cdot R(Q)_i
	-\frac{1}{4}\varepsilon_{ijk}\gamma \cdot R(V){}^j{}_l \chi^{kl} \nn\\
	&\quad - \frac{i}{3} \gamma \cdot R(A) \zeta_i -\frac{1}{12}\gamma \cdot R(V){}^j_i\zeta_j 
	+ 2\varepsilon_{ijk} R(S)^j \cdot T^k    -\frac{1}{96} \varepsilon_{ikl} D^l\;_j \chi^{jk}  -\frac{1}{96}D_i\;^j \zeta_j
	\nn\\
	&\quad  + \frac{1}{16} E^j \slashed{D} \chi_{ij} + \frac{1}{16} \varepsilon_{ijk} E^j \slashed{D} \zeta^k -{ \frac{1}{48}} \varepsilon_{ijk} \gamma \cdot T^j\slashed{D}\zeta^k +\frac{1}{16} \gamma \cdot T^j \slashed{D} \chi_{ij} + 2D_aD^c T_{cbi} \gamma^{ab} \Lambda_R 		\nn \\
	&\quad  -\frac{1}{2}D^aD_a E_i \Lambda_R + \frac{1}{4} \varepsilon_{ijk} \gamma^{ab} T_{ca}{}^k T^c\;_b^j \slashed{D}\Lambda_L 
	+ \varepsilon_{ijk} \gamma^c D_a T_{bc}^j T^{abk} \Lambda_L- \frac{1}{3} \varepsilon_{ijk}\gamma^a T_{ab}{}^j D_c T^{bck} \Lambda_L  \nn \\
	&\quad +\frac{1}{12}\varepsilon_{ijk}T^k \cdot \slashed{D} T^j \Lambda_L + \frac{1}{16} \varepsilon_{ijk} E^j \slashed{D}E^k \Lambda_L + \frac{1}{24}\varepsilon_{ijk} \gamma \cdot \slashed{D} T^k E^j \Lambda_L + {\frac{1}{24}}\varepsilon_{ijk}\gamma \cdot T^k \slashed{D} E^j \Lambda_L
	\nonumber \\
	&\quad - R(V){}^j_i \cdot T_j \Lambda_R + 2i R(A) \cdot T_i \Lambda_R - \frac{i}{2} \gamma \cdot R(A) E_i \Lambda_R 
	-\frac{1}{6}E_i R(Q)_j \cdot T^j  - \frac{1}{128} \varepsilon_{ijk} E_l E^j \chi^{kl}\nn\\
	&\quad + \frac{1}{128}\varepsilon_{ijk}E_l \gamma \cdot T^j \chi^{kl}
	- \frac{1}{384} \zeta_i E_jE^j - \frac{1}{384}E_i E^j\zeta_j  - \frac{1}{128}\gamma \cdot T^j E_j \zeta_i - \frac{1}{1152} \gamma \cdot T^j \zeta_j E_i
	\nn\\
	&\quad  -\frac{1}{96} E_j D^j\;_i \Lambda_R   - \frac{1}{64} E_iE_j E^j \Lambda_R - \frac{3}{16} T_i \cdot T_j \gamma \cdot T^j \Lambda_R  - \frac{7}{48}T_i \cdot T_j E^j \Lambda_R 
	\nn \\
	&\quad- \frac{7}{192} E_i E_j \gamma \cdot T^j \Lambda_R 
	\end{align}
	The { purely bosonic part of the } singlet $Y$ appearing in the right-supersymmetry transformation of composite $\mathcal{N}_i$ reads as follows.
	\begin{align}
	Y &= -24R(M)^{abcd}R(M)^+_{abcd}-48R(V)^+{}_i{}^j\cdot R(V)^+{}_i{}^j +144 R(A) \cdot R^+(A) +3R(V){}^j_i\cdot T^iE_j		\nn \\
	&\quad-3R(V){}^j_i\cdot T_jE^i -6iR(A)\cdot T^jE_j+ 6iR(A)\cdot T_jE^j -24 T^{abi} D_a D^c T_{bci} -\frac{3}{2} E^i D^a D_a E_i		\nn\\
	&\quad-\frac{1}{48}D^i\;_j D^j\;_i -\frac{13}{96}T^i\cdot T^jE_iE_j-\frac{19}{96}T_i\cdot T_jE^iE^j+ \frac{1}{384}E_iE^iE_jE^j - \frac{7}{8} \Big(T^i \cdot T^j \Big) \Big(T_i \cdot T_j \Big)
	\end{align}
	The composite $\mathcal{L}$ is then given by evaluation \eqref{3.39} and is presented in section-\ref{sec-actions}.
	
	\section{Composites for the Weyl square action}\label{appendix-weylsquare}
	In this section, we will present the composites that contribute to the Weyl square density. As we discussed in section-\ref{sec-actions}, the embedding to obtain the Weyl square action is given as, 
	\begin{align}\label{cdef2}
	C^{ij}=i{}\; T^i\cdot T^j + i\; \bar{\Lambda}_R\chi^{ij}+ \frac{i}{4}E^iE^j.
	\end{align}
	From \eqref{csusy}, we see that the composites $\rho^i{}_j$ and $\hat{\rho}^i$ differ from the previous section by a factor of $i$ as given below.
	\begin{align}\label{rhodef2}
	\hat{\rho}^i&= - 4i\;\slashed{D} \gamma \cdot T^i \Lambda_R - 4i\;T^{abi} \slashed{D} \gamma_{ab} \Lambda_R
	- 4i\; E^i \slashed{D} \Lambda_R -4i\; \slashed{D} E^i \Lambda_R  \nonumber\\
	&\quad - \frac{4i}{3} \bar{\Lambda}_R\Lambda_R \zeta^i   - 8i\; \bar{\Lambda}_R \Lambda_R E^i \Lambda_L, \\
	\rho^i{}_l&= \frac{3i}{8}\varepsilon_{jkl}\gamma\cdot T^k\chi^{ij}-\frac{i}{8}(\gamma\cdot T^i\zeta_l-\frac{1}{3}\delta^i_l\gamma\cdot T^k\zeta_k)+ 12i\;(T^{abi}R(Q)_{abl}-\frac{1}{3}\delta^i_lT^{abk}R(Q)_{abk})\nonumber\\
	&\quad+ \frac{3i}{8}\varepsilon_{jkl}E^k\chi^{ij}+\frac{3i}{8}(E^i\zeta_l-\frac{1}{3}\delta^i_lE^k\zeta_k)  + 12i\;\gamma\cdot R(V){}^i{}_l\Lambda_R-\frac{i}{2}D^l_i\Lambda_R\nonumber\\
	&\quad-\frac{3i}{4}(\gamma\cdot T^i\Lambda_RE_l-\frac{1}{3}\delta^i_l\gamma\cdot T^kE_k\Lambda_R).
	\end{align}
	The composite $F_{ij}$ and $G_{ab}{}^i$ obtained from the right-supersymmetry transformation of $\rho^i{}_j$ (\ref{right_susy_rho}), are also rescaled by a factor of $i$ as,
	\begin{align}
	F_{ij}&= 96i\;\bar{R}(Q)^{ab}_{i}R(Q)_{abj}-\frac{3i}{4}\varepsilon_{ikl}\varepsilon_{jmn}\bar{\chi}^{km}\chi^{ln}-12i\;\bar{\Lambda}_{R}\slashed{D}\chi_{ij}+\frac{i}{2}\bar{\zeta}_{i}\zeta_{j}  +2i\;\bar{\Lambda}_{R}\zeta_{(i}E_{j)}\nn \\
	&\quad  -96i\;\varepsilon_{kl(i}R(V){}^k{}_{j)}\cdot T^l -i\;\varepsilon_{kl(i} E^k D^l{}_{j)} + 12i\; {T_i\cdot T_j \bar{\Lambda }_R\Lambda_R} + 3i\; {E_iE_j\bar{\Lambda }_R\Lambda_R}\\
	G_{ab}^{i}&=32i\;\bar{R}(Q)_{abk}\chi^{ik}-\frac{2i}{3}\bar{\zeta}_{k}\gamma_{ab}\chi^{ik}-\frac{32i}{3}\varepsilon^{ikl}\bar{\zeta}_{k}R(Q)_{abl}-\frac{2i}{9}\varepsilon^{ikl}\bar{\zeta}_{l}\gamma_{ab}\zeta_{k}-\frac{16i}{3}\bar{\Lambda}_{R}\gamma_{ab}\slashed{D}\zeta^{i}\nonumber \\
	&\quad +64i\;\varepsilon^{ijk}\bar{R}(Q)^{cd}_{j}\gamma_{ab}R(Q)_{cdk}+256i\;\bar{\Lambda}_{R}R(S)_{ab}^{+i} + 32i\;\varepsilon^{ijk}\bar{\Lambda}_{R}R(Q)_{abj}E_{k}\nn\\
	&\quad  - 24i\; E^i \bar{\Lambda}_L \slashed{D} \gamma_{ab}\Lambda_R  - 32i\; E^i \bar{\Lambda}_R   \gamma_{ab} \slashed{D} \Lambda_L - 8i\; \bar{\Lambda}_L \slashed{D} \gamma_{ab} E^i \Lambda_R +64i\;\bar{\Lambda}_L\slashed{D}T_{ab}^i\Lambda_R\nn \\
	&\quad +32i\;\bar{\Lambda}_L\gamma^d\gamma_{ab}\Lambda_RD^cT_{cd}^i  -\frac{128i}{3}T_{ab}^i\bar{\Lambda }_L\slashed{D}\Lambda_R-\frac{256i}{3}T_{ab}^i\bar{\Lambda}_R\slashed{D}\Lambda_L 
	-8i\;\bar{\Lambda}_L\gamma^c\gamma\cdot T^i \gamma_{ab}D_c\Lambda_R
	\nn \\
	&\quad- 64i\; R(V)_{ab}{}^i{}_j E^j + 128\; E^i R(A)_{ab} + 256i\; T_{c[a}{}^nR(V)^c{}_{b]}{}^i{}_n+256i\;T_{c[a}^iR(A)^c{}_{b]} \nn \\
	&\quad+128i\;T^{cdi}R(M)_{cdab}-\frac{8i}{3}T_{ab}^jD_j^i  + \frac{8i}{3}T_{ab}{}^i E^j E_j   +  16i\; T_{c[a}{}^i T^c{}_{b]}{}^m E_m 
	\end{align}
	From \eqref{H-composite}, we see that the second term has a factor of $-i$ due to the presence of $C_{ij}$. Therefore the composite $\theta_i$ that appears in the transformation of $K_{ij}$ of  Weyl square density differs non-trivially from that of Pontryagin density. It is given as follows.
	\begin{align}
	\theta_i&=-8i\;\gamma^{cd}R(Q)^{ab}{}_iR(M)_{abcd} +32i\;R(V){}^j_i\cdot R(Q)_j+16R(A)\cdot R(Q)_i
	+{i}\varepsilon_{ikl}\gamma\cdot R(V){}^l{}_j\chi^{jk} \nn \\
	&\quad+\frac{4}{3}\gamma\cdot R(A) \zeta_i-\frac{i}{3}\gamma\cdot  R(V){}^j{}_i\zeta_j+8i\;\varepsilon_{ijk}R(S)^j\cdot T^k-\frac{i}{24}\varepsilon_{ikl}\chi^{jk}D^l_j -{\frac{i}{24}}D_i^j\zeta_j
	\nn \\
	&\quad +\frac{i}{4} E^j \slashed{D} \chi_{ij}  +  \frac{i}{4} \varepsilon_{ijk} E^j \slashed{D} \zeta^{k} +\frac{i}{4}\gamma\cdot T^j\slashed{D}\chi_{ij} +\frac{i}{12}\varepsilon_{ijk}\gamma\cdot T^k\slashed{D}\zeta^j
	- 2i\; D^a D_a E_i \Lambda_R\nn\\
	&\quad +8i\;D_aD^cT_{cbi}\gamma^{ab}\Lambda_R +i\;\varepsilon_{ijk}\gamma^{ab}T_{ca}^kT^c{}_b{}^j\slashed{D}\Lambda_L+\frac{3i}{128}\varepsilon_{ijk}E_l\gamma\cdot T^j \chi^{kl} - \frac{5i}{128} \varepsilon_{ijk} E_l E^j \chi^{kl}\nn\\
	&\quad- \frac{5i}{96}  E_j D^j\;_i  \Lambda_R-\frac{i}{384}\gamma\cdot T^j\zeta_jE_i-\frac{13i}{384}\gamma\cdot T^jE_j\zeta_i -\frac{i}{384}\zeta_i E_j E^j -  \frac{5i}{384} E_i E^j \zeta_{j}\nn\\
	&\quad +4i\;\varepsilon_{ijk}\gamma^cD_aT_{bc}{}^jT^{abk}\Lambda_L+  \frac{i}{4} \varepsilon_{ijk} E^j \slashed{D} E^{k} \Lambda_L +\frac{i}{4}\varepsilon_{ijk}\gamma\cdot\slashed{D}T^k\Lambda_LE^j +\frac{i}{4}\varepsilon_{ijk}\gamma\cdot T^k\slashed{D}E^j\Lambda_L  \nn\\
	&\quad-4i\;R(V){}^j_i\cdot T_j\Lambda_R- 8 R(A)\cdot T_i\Lambda_R +\frac{i}{4} \gamma \cdot R(V){}^j{}_i E_j \Lambda_R + 2 E_i \gamma \cdot R(A) \Lambda_R \nn\\
	&\quad +\frac{i}{4}R(Q)_i\cdot T^jE_j
	-\frac{3i}{4}E_iR(Q)_j\cdot T^j-\frac{3i}{8} T_i\cdot T_j E^j\Lambda_R-\frac{3i}{8} T_i\cdot T_j \gamma\cdot T^j\Lambda_R
	\nn\\
	&\quad-\frac{i}{16}E_iE_j\gamma\cdot T^j\Lambda_R -  \frac{i}{96} E_i E^j E_j \Lambda_R - \frac{i}{3}\bar{\Lambda}_L\slashed{D}\Lambda_R\zeta_i  +\frac{i}{4} \gamma_{ab}\zeta_i\bar{\Lambda}_L\slashed{D}\gamma^{ab}\Lambda_R  
	+ \frac{2i}{3}\bar{\Lambda}_R\slashed{D}\Lambda_L \zeta_i\nn\\
	&\quad+ \frac{i}{6} \gamma_{ab} \zeta_i\bar{\Lambda}_R \gamma^{ab} \slashed{ D}\Lambda_L +\frac{i}{3} \bar{\Lambda}_R \gamma^{ab}D_b \zeta_i \gamma_a \Lambda_L-\frac{i}{3}\bar{\Lambda}_R D_a \zeta_i \gamma^a\Lambda_L - \frac{i}{12}\bar{\zeta}^j\chi_{ij} \Lambda_R  \nn\\
	&\quad + 8i\;\bar{\Lambda}_R \gamma^a D^b\Lambda_L R(Q)_{abi} - 8i\; \bar{\Lambda}_R \gamma^a \Lambda_L D^b R(Q)_{abi} -\frac{i}{8}\bar{\Lambda}_L\chi_{ij}\gamma\cdot T^j\Lambda_R -\frac{i}{8}\bar{\Lambda}_L\chi_{ij}E^j\Lambda_R  \nn\\
	&\quad +\frac{4i}{3} \bar{\Lambda}_R \Lambda_R E_i \slashed{D}\Lambda_L + \frac{i}{3}E_i\bar{\Lambda}_L\slashed{D}\Lambda_R \Lambda_R+ \frac{i}{4}E_i\gamma^{ab}\Lambda_R\bar{\Lambda}_L\slashed{D}\gamma_{ab}\Lambda_R  \nonumber\\
	&\quad + i \Lambda_R \bar{\Lambda}_L \gamma \cdot T_i \slashed{D} \Lambda_R  - { \frac{i}{4} \gamma^{ab}\Lambda_R \bar{\Lambda}_L \gamma \cdot T_i  \slashed{D} \gamma_{ab} \Lambda_R} + {\frac{i}{6}}\varepsilon_{ijk}\Lambda_R\bar{\Lambda}_L\zeta^kE^j
	\nn \\
	&\quad+ 2i\; \bar{\Lambda}_R \Lambda_R \slashed{D} \gamma \cdot T_i \Lambda_L    - \frac{i}{4} \varepsilon_{ijk} E^k \gamma \cdot T^l \Lambda_R \bar{\Lambda}_L \Lambda_L
	+\frac{i}{6}\bar{\Lambda}_R\Lambda_R\bar{\Lambda}_L\Lambda_L\zeta_i 
	\end{align}
	The composite $E_{ai}$ defined in \eqref{Eadefinition}  acquires a factor of $i$ relative to its expression in Pontryagin density and is given by,
	\begin{align}
	E_{ai}&=-64i\varepsilon_{ijk}T_{ab}{}^jD^cT^b{}_c{}^k+16i\varepsilon_{ijk}T^k\cdot D_aT^j  - 16i \varepsilon_{ijk} E^j{D^b} T_{ab}{}^k  -16i \varepsilon_{ijk}  T^{abk} D_b E^j \nn\\
	&\quad - \frac{16i}{3} D^b \bar{\Lambda}_R \gamma_{ab} \zeta_i -\frac{16i}{3}\bar{\Lambda}_R\gamma_{ab}D^b\zeta_i -256\overline{D^b\Lambda_R} R(Q)_{abi} -256i\bar{\Lambda}_RD^bR(Q)_{abi} \nn\\
	&\quad + i \varepsilon_{ijk} \bar{ \Lambda}_R\gamma_a \zeta^j E^k  +i \bar{\Lambda}_R \gamma_a \chi_{ij} E^j
	-32i\bar{\Lambda}_R\Lambda_RD^bT_{bai}+64i\bar{\Lambda}_RD^b\Lambda_RT_{abi} 
	\nn\\
	&\quad - 16i\bar{\Lambda}_RD_a\Lambda_RE_i - 8i\bar{\Lambda}_R\Lambda_RD_aE_i+4i\varepsilon_{ijk}\bar{\Lambda}_R\gamma^b\zeta^jT_{ab}{}^k+4i\bar{\Lambda}_R\gamma^b\chi_{ij}T_{ab}{}^j\nn\\
	&\quad+16i\varepsilon_{ijk}\bar{\Lambda}_R\gamma^b\Lambda_LT_{ab}{}^kE^j+16i\varepsilon_{ijk}\bar{\Lambda}_R\gamma^b\Lambda_LT_{bc}{}^jT_a{}^{ck} -\frac{16i}{3}\bar{\Lambda}_R\Lambda_R\bar{\Lambda}_L\gamma_a\zeta_i
	\end{align}
	
	It can be seen from \eqref{tildezetas} that the composite $\tilde{\theta}_i$ also has non trivial differences compared to that of the Pontryagin density, as given below. 
	\begin{align} 
	\tilde{\theta}_i&= - \frac{3i}{8}\varepsilon_{ikl}  E_j \gamma \cdot T^l  \chi^{jk} - \frac{3i}{8} \varepsilon_{ikl} E^l  E_j \chi^{jk} + \frac{i}{2} E_j D^j\;_i \Lambda_R - \frac{i}{24} \gamma \cdot T^j \zeta_jE_i
	\nonumber\\
	&\quad  + \frac{i}{8} \gamma \cdot T^j E_j \zeta_i  -\frac{3i}{8}\zeta_i E^j E_j + \frac{i}{8} E_i E^j \zeta_j -{16i \varepsilon_{ijk} E^j {D}^b T_{ab}{}^k \gamma^a \Lambda_L}
	\nonumber\\
	&\quad  -64i\varepsilon_{ijk}\gamma^aT_{ab}{}^jD^cT^b{}_c{}^k\Lambda_L+16i\varepsilon_{ijk}T^k\cdot\slashed{D}T^j\Lambda_L  -16i \varepsilon_{ijk} \gamma_a T^{abk} D_b E^j \Lambda_L 
	\nonumber \\
	&\quad  - 12i \gamma \cdot R(V){}^j_i E_j \Lambda_R -12i  R(Q)_{i}\cdot T^{ j}  E_j + 4i E_i R(Q)_{ j} \cdot T^{ j} 
	\nonumber \\
	&\quad + \frac{i}{2}  E_i E_j \gamma \cdot T^j\Lambda_R  -{\frac{1}{2}  E_i E_j E^j\Lambda_R - 2i T_i \cdot T_j E^j \Lambda_R  }  + 8i\bar{\Lambda}_L \slashed{D}\Lambda_R \zeta_i 	\nn \\
	&\quad+ \frac{8i}{3}\bar{\Lambda}_L\gamma^a D^b \Lambda_R \gamma_{ab}\zeta_i  -\frac{16i}{3}\bar{\Lambda}_R\gamma_{ab}D^b\zeta_i\gamma^a\Lambda_L  + 256i\overline{D^b\Lambda}_R\gamma^a\Lambda_LR(Q)_{abi} 
	\nn \\
	&\quad  +256i\bar{\Lambda}_R\gamma^a\Lambda_LD^bR(Q)_{abi} + i  \varepsilon_{ijk} \bar{ \Lambda}_R\gamma_a \zeta^j E^k \gamma^a \Lambda_L +64i\gamma^a\bar{\Lambda}_RD^b\Lambda_RT_{abi}\Lambda_L 
	 \nn \\ 
	&\quad +32i\gamma^aT_{abi}D^b\Lambda_L\bar{\Lambda}_R\Lambda_R - 16i\bar{\Lambda}_RD_a\Lambda_RE_i\gamma^a\Lambda_L-8iE_i\slashed{D}\Lambda_L\bar{\Lambda}_R\Lambda_R  \nn\\
	&\quad+ 4i\varepsilon_{ijk}\bar{\Lambda}_R\gamma^b\zeta^jT_{ab}{}^k\gamma^a\Lambda_L+4i\bar{\Lambda}_R\gamma^b\chi_{ij}T_{ab}{}^j\gamma^a\Lambda_L 
	-16i\bar{\Lambda}_R\Lambda_R\slashed{D}\gamma\cdot T_i\Lambda_L \nn\\
	&\quad+ 16i\varepsilon_{ijk}\bar{\Lambda}_R\gamma^b\Lambda_L\gamma^aT_{ab}{}^kE^j\Lambda_L+16i\varepsilon_{ijk}\bar{\Lambda}_R\gamma^b\Lambda_LT_{bc}{}^jT_a{}^{ck}\gamma^a\Lambda_L+8i\;\bar{\Lambda}_R\Lambda_R\bar{\Lambda}_L\Lambda_L\zeta_i 
	\end{align}
	The composite $\mathcal{N}_i$ is given as follows ,
	\begin{align}
	\mathcal{N}_i &= 2i\;\gamma^{cd} R(Q)^{ab}_iR(M)_{abcd}  + 8i\; R(V){}^j{}_i \cdot R(Q)_j + 4 R(A)\cdot R(Q)_i
	\nn\\
	&\quad -\frac{i}{4}\varepsilon_{ijk}\gamma \cdot R(V){}^j{}_l \chi^{kl}  + \frac{1}{3} \gamma \cdot R(A) \zeta_i -\frac{i}{12}\gamma \cdot R(V){}^j{}_i\zeta_j 
	\nonumber \\
	&\quad+{2i}\;\varepsilon_{ijk} R(S)^j \cdot T^k   -\frac{i}{96} \varepsilon_{ikl} D^l\;_j \chi^{jk} -\frac{i}{96}D_i\;^j \zeta_j
	\nn \\
	&\quad  + \frac{i}{16} E^j \slashed{D} \chi_{ij} + \frac{i}{16} \varepsilon_{ijk} E^j \slashed{D} \zeta^k - {\frac{i}{48}} \varepsilon_{ijk} \gamma \cdot T^j\slashed{D}\zeta^k +{\frac{i}{16}\gamma \cdot T^j \slashed{D}\chi_{ij}}
	\nn\\
	&\quad + 2i\;D_aD^c T_{cbi} \gamma^{ab} \Lambda_R -\frac{i}{2}D^aD_a E_i \Lambda_R 
	+ \frac{i}{4} \varepsilon_{ijk} \gamma^{ab} T_{ca}{}^k T^c\;_b^j \slashed{D}\Lambda_L 
	\nonumber \\
	&\quad  +{i}\; \varepsilon_{ijk} \gamma^c D_a T_{bc}^j T^{abk} \Lambda_L - \frac{i}{3} \varepsilon_{ijk}\gamma^a T_{ab}{}^j D_c T^{bck} \Lambda_L +\frac{i}{12}\varepsilon_{ijk}T^k \cdot \slashed{D} T^j \Lambda_L 
	\nn \\
	&\quad + \frac{i}{16} \varepsilon_{ijk} E^j \slashed{D}E^k \Lambda_L + \frac{i}{24}\varepsilon_{ijk} \gamma \cdot \slashed{D} T^k E^j \Lambda_L  +{\frac{i}{24}}\varepsilon_{ijk}\gamma \cdot T^k \slashed{D} E^j \Lambda_L
	\nonumber \\
	&\quad -i\; R(V){}^j{}_i \cdot T_j \Lambda_R - 2 R(A) \cdot T_i \Lambda_R + \frac{1}{2} \gamma \cdot R(A) E_i \Lambda_R 
	\nn\\
	&\quad  - \frac{i}{6} E_i R(Q)_j \cdot T^j  -\frac{i}{96} E_j D^j\;_i \Lambda_R - \frac{i}{128} \varepsilon_{ijk} E_l E^j \chi^{kl}  + \frac{i}{128}\varepsilon_{ijk}E_l \gamma \cdot T^j \chi^{kl} 
	\nonumber \\
	&\quad- \frac{i}{384} \zeta_i E_jE^j - \frac{i}{384}E_i E^j\zeta_j- \frac{i}{128}\gamma \cdot T^j E_j \zeta_i - \frac{i}{1152} \gamma \cdot T^j \zeta_j E_i 
	\nonumber \\
	&\quad  - \frac{i}{192} E_iE_j E^j \Lambda_R  -{ \frac{i}{16}} T_i \cdot T_j \gamma \cdot T^j \Lambda_R - {\frac{5i}{48}} {T_i \cdot T_j E^j \Lambda_R} -{ \frac{i}{192}}{\gamma \cdot T^j \Lambda_R E_i E_j}
	\nonumber \\
	&\quad      - \frac{2i}{3} \bar{\Lambda}_R \gamma^a \Lambda_L D^b R(Q)_{abi}  + 2i\; \Lambda_R \gamma^a D^b \Lambda_L R(Q)_{abi} +\frac{4i}{3}\overline{D^b\Lambda}_R\gamma^a\Lambda_LR(Q)_{abi}
	\nonumber \\
	&\quad -\frac{i}{48}\bar{\zeta}^j \chi_{ij} \Lambda_R + \frac{i}{6} \bar{\Lambda}_R \slashed{D} \Lambda_L \zeta_i + \frac{i}{24}\gamma^{ab} \zeta_i \bar{\Lambda}_R \gamma_{ab} \slashed{D} \Lambda_L  - \frac{i}{12} \bar{\Lambda}_R D_a \zeta_i \gamma^a \Lambda_L  
	\nonumber \\
	&\quad+ \frac{i}{18} \bar{\Lambda}_R \gamma^{ab} D_b \zeta_i \gamma_a \Lambda_L - \frac{i}{24} \bar{\Lambda}_L \slashed{D}\Lambda_R \zeta_i + \frac{17i}{288} \gamma^{ab} \zeta_i \bar{\Lambda}_L \slashed{D}\gamma_{ab} \Lambda_R 
	\nn\\ 
	&\quad -{ \frac{i}{32}}\bar{\Lambda}_L \chi_{ij}E^j \Lambda_R + \frac{i}{96} \bar{\Lambda}_L \chi_{ij} \gamma \cdot T^j \Lambda_R + { \frac{i}{32}}\varepsilon_{ijk} \Lambda_R \bar{\Lambda}_L \zeta^k E^j 
	\nn \\
	&\quad +\frac{i}{48} \varepsilon_{ijk} \bar{\Lambda}_R \gamma^b \zeta^j T_{ab}{}^k \gamma^a \Lambda_L  + \frac{5i}{12} \bar{\Lambda}_R \Lambda_R \slashed{D} \gamma \cdot T_i \Lambda_L  + { \frac{7i}{24}} \bar{\Lambda}_R \Lambda_R E_i \slashed{D}\Lambda_L 
	\nn \\
	&\quad + \frac{i}{24}E_i \Lambda_R \bar{\Lambda}_L \slashed{D}\Lambda_R + {\frac{5i}{96}} E_i \gamma^{ab} \Lambda_R \bar{\Lambda}_L \slashed{D} \gamma_{ab}\Lambda_R + \frac{7i}{24} \Lambda_R \bar{\Lambda}_L \gamma \cdot T_i \slashed{D}\Lambda_R 
	\nn \\ 
	&\quad - \frac{i}{16} \gamma^{ab} \Lambda_R \bar{\Lambda}_L \gamma \cdot T_i \slashed{D}\gamma_{ab}\Lambda_R -\frac{i}{6}\bar{\Lambda}_L\gamma^{(a}D^{b)}\Lambda_R T^c{}_{ai} \gamma_{bc} \Lambda_R  -  \frac{i}{24} { \bar{\Lambda}_R \Lambda_R \slashed{D} \gamma^{ab}\Lambda_L T_{abi}}
	\nn \\
	&\quad + \frac{i}{12}\varepsilon_{ijk} \bar{\Lambda}_R\gamma^b \Lambda_L T_{bc}^j T_{a}{}^{ck} \gamma^a \Lambda_L  \; + \frac{i}{48} \varepsilon_{ijk} E^k \gamma \cdot T^j \Lambda_R \bar{\Lambda}_L \Lambda_L
	\nn \\
	&\quad +\frac{i}{12}\bar{\Lambda}_R\Lambda_R \bar{\Lambda}_L\Lambda_L \zeta_i 
	\end{align}
	The singlet $Y$ appearing in the Lagrangian is obtained from the right-supersymmetry transformation of $N_i$.
	\begin{align}
	Y&=-24iR(M)^{abcd}R(M)^+_{abcd} -48iR(V)^+{}_i{}^j\cdot R(V)^+{}_i{}^j +144i R(A) \cdot R^+(A)
	\nn \\
	&\quad + 3iR(V){}^j{}_i\cdot T^iE_j - 3iR(V){}^j_i\cdot T_jE^i +6R(A)\cdot T^jE_j-6R(A)\cdot T_jE^j\nn \\
	&\quad - 24i T^{abi} D_a D^c T_{bci} -\frac{3i}{2} E^i D^a D_a E_i  -\frac{i}{48}D^i\;_j D^j\;_i 
	-\frac{7i}{96}T^i\cdot T^jE_iE_j \nonumber \\
	&\quad -\frac{17i}{96}T_i\cdot T_jE^iE^j + \frac{i}{128}E_iE^iE_jE^j  - \frac{5i}{8} \Big(T^i \cdot T^j \Big) \Big(T_i \cdot T_j \Big)  -48i\overline{R(S)}^i\cdot R(Q)_i
	\nn \\
	&\quad + \frac{3i}{8}\bar{\chi}^{ij}\slashed{D} \chi_{ij}+\frac{i}{4}\bar{\zeta}_i\slashed{D}\zeta_i +12\bar{\Lambda}_R\slashed{D}D^2\Lambda_L+12\bar{\Lambda}_RD^2\slashed{D}\Lambda_L -\frac{i}{16} \bar{\chi}^{ij} \zeta_i E_j \nn\\
	&\quad-\frac{i}{16} \bar{\chi}_{ij}\zeta^i E^j  +\frac{i}{4}\bar{\zeta}_i\slashed{D}\Lambda_LE^i-\frac{i}{4}\bar{\Lambda}_L\slashed{D}E^j\zeta_j + \frac{i}{4}\bar{\Lambda}_R \slashed{D}\zeta^i E_i - {\frac{5i}{12}} \bar{\zeta}_i \gamma \cdot T^i \slashed{D} \Lambda_L\nn\\
	&\quad +{ \frac{i}{24}}\bar{\zeta}_i \gamma \cdot \slashed{D} T^i \Lambda_L + {\frac{13i}{24}} \bar{\zeta}^i \gamma \cdot \slashed{D} T_i \Lambda_R  +{ \frac{i}{4}} \bar{\Lambda}_R \gamma^a \gamma \cdot T_i D_a \zeta^i - {\frac{i}{6} }\bar{\Lambda}_L \gamma^a \gamma \cdot T^i D_a \zeta_i 
	\nn\\
	&\quad + {18i} \bar{\Lambda}_R \slashed{D} R(Q)^i \cdot T_i - 2i\bar{\Lambda}_L\slashed{D}R(Q)_{i} \cdot T^{i} + {{6i}}\bar{\Lambda}_R \slashed{D} T_i \cdot R(Q)^i - {8i}\bar{\Lambda}_L\slashed{D}T^i\cdot R(Q)_i  \nn\\
	&\quad +14iT^i\cdot \overline{R(Q)}_i\slashed{D}\Lambda_L -{48}\bar{\Lambda}_L\gamma^a\gamma\cdot R(A)D_a\Lambda_R
	+ {9}\bar{\Lambda}_R\gamma^a\gamma\cdot R(A)D_a\Lambda_L\nn \\
	&\quad - {39}\bar{\Lambda}_R\gamma\cdot R(A)\slashed{D}\Lambda_L  - {6}\bar{\Lambda}_R \gamma^a \Lambda_L D^b R(A)_{ab} -\frac{7i}{64} \bar{\Lambda}_L \chi_{ij}E^i E^j -\frac{5i}{64}\bar{\Lambda}_R\chi^{ij}E_iE_j \nn \\
	&\quad - {\frac{17i}{48}}\Lambda_L \chi_{ij} T^i \cdot T^j -{\frac{31i}{48}}\bar{\Lambda}_R\chi^{ij}T_i\cdot T_j
	-\frac{7i}{16}\bar{\Lambda}_R\slashed{D}\Lambda_LE^iE_i+ \frac{3i}{16} \bar{\Lambda}_L \slashed{D} \Lambda_R E_i E^i \nonumber\\
	&\quad +\frac{41i}{48} \bar{\Lambda}_L \gamma \cdot T_i \slashed{D} \Lambda_R E^i +\frac{5i}{48} \bar{\Lambda}_R\gamma\cdot T^i\slashed{D}\Lambda_LE_i+{\frac{33i}{16}}\bar{\Lambda}_R \slashed{D}\gamma^{ab}\Lambda_LT_{abi}E^i
	\nn \\
	&\quad  +{\frac{11i}{16}}  \bar{\Lambda}_L \slashed{D}\gamma_{ab}\Lambda_R T^{abi} E_i  +\frac{19i}{3}\overline{D^b\Lambda_L}\gamma^c\Lambda_RT^{abj}T_{acj} -\frac{23i}{3}\bar{ \Lambda}_L\gamma^cD_b\Lambda_RT_{acj}T^{abj} \nonumber\\
	&\quad +{\frac{21i}{16}}\bar{\Lambda}_L\gamma\cdot\slashed{D}T_i\Lambda_RE^i - {\frac{i}{16}}\bar{\Lambda}_L\slashed{D}\gamma\cdot T^j\Lambda_RE_j+ \frac{13i}{3}\bar{\Lambda}_L\gamma^c\Lambda_R D_bT^{abj}T_{acj}
	\nonumber\\
	&\quad - \frac{17i}{3}\bar{ \Lambda}_L\gamma^c\Lambda_RT^{abj}D_bT_{acj} +\frac{5i}{12} \bar{\Lambda}_R\Lambda_R\varepsilon^{ijk}T^{ab}{}_iT^c{}_a{}_jT_{bck} -\frac{5i}{12}\varepsilon_{ijk} \bar{\Lambda}_L \Lambda_L T^{abi} T^c{}_{a}^{j} T_{bc}^{k} 
	 \nn\\
	&\quad - \frac{11i}{16} \bar{\Lambda}_L \slashed{D}E_i \Lambda_R E^i - \frac{i}{16}  \bar{\Lambda}_L \slashed{D}E^i \Lambda_R E_i - \frac{31i}{48}\bar{\Lambda}_R \gamma \cdot T^i \slashed{D} E_i \Lambda_L
	\nn \\
	&\quad + {\frac{13i}{48}}\bar{\Lambda}_L \gamma \cdot T_i \slashed{D} E^i \Lambda_R +{\frac{i}{6}}\bar{\Lambda}_L\zeta^i\bar{\Lambda }_R\zeta_i + 30 i  \bar{\Lambda}_R \Lambda_R \bar{\Lambda}_L D^aD_a \Lambda_L +  { 6 i } \bar{\Lambda}_L \Lambda_L \bar{\Lambda}_R D^aD_a \Lambda_R 
	\nn \\
	& \quad   - 2 i  \bar{\Lambda}_R\Lambda_R \overline{D_a \Lambda}_L  {D }^a\Lambda_L -   12 i  \bar{\Lambda}_R\Lambda_R \overline{D_a \Lambda}_L \gamma^{ab}  {D }_b\Lambda_L  - 26 i  \bar{\Lambda}_L\Lambda_L \overline{D_a \Lambda}_R  {D }^a\Lambda_R  
	\nonumber\\
	&\quad + { 8 i  \bar{\Lambda}_L\Lambda_L \overline{D_a \Lambda}_R \gamma^{ab}  {D }_b\Lambda_R }  -{ 8 i }\bar{ \Lambda}_RD^a\Lambda_R \bar{\Lambda}_LD_a\Lambda_L  - { 40 i} \bar{ \Lambda}_RD_a\Lambda_R \bar{ \Lambda}_L\gamma^{ab}D_b\Lambda_L
	\nn\\
	& \quad  + {16 i } \bar{ \Lambda}_R\gamma^{ab}D_b\Lambda_R \bar{ \Lambda}_L \gamma_{ac} D^c\Lambda_L + {\frac{7i}{24}}\bar{\Lambda}_L\Lambda_L\bar{\Lambda}_R\zeta_jE^j + {\frac{5i}{24}}\bar{\Lambda}_R \Lambda_R \bar{\Lambda}_L E_i \zeta^i
	\nn \\
	&\quad  + {\frac{7i}{36}}\bar{\Lambda}_L\Lambda_L\bar{\Lambda}_R\gamma\cdot T^i\zeta_i+ {\frac{19i}{72}}\bar{\Lambda}_R \Lambda_R \bar{\Lambda}_L \gamma \cdot T_i \zeta^i  +\frac{11i}{3} \bar{\Lambda}_L\Lambda_L\bar{\Lambda}_RR(Q)_i\cdot T^i 
	\nonumber\\
	&\quad +\frac{25i}{3} \bar{\Lambda}_R\Lambda_R\bar{\Lambda}_LR(Q)^i\cdot T_i 
	\end{align}	

	\bibliography{references}
	\bibliographystyle{jhep}

\end{document}